\documentclass[a4paper,11pt,twoside]{article}
\usepackage[utf8]{inputenc}

\usepackage[T1]{fontenc}
\usepackage{charter}
\usepackage[expert]{mathdesign}

\usepackage{titlesec} 
\usepackage{cite}
\usepackage{amsmath}
\usepackage{amssymb}
\usepackage{amsfonts}
\usepackage{graphicx}
\usepackage{sectsty}
\usepackage[DIV=11,BCOR=2mm,headinclude=true,footinclude=false]{typearea}
\usepackage{tikz}
\usepackage{booktabs}
\usepackage{mathrsfs}
\usepackage{todonotes}
\usepackage{acronym}
\usepackage{dsfont}
\usepackage[ruled,vlined,linesnumbered]{algorithm2e}
\usepackage[colorlinks=true]{hyperref}
\usepackage{cleveref}
\usepackage{subcaption}
\usepackage{rotating}
\usepackage{mathtools}
\usepackage{listings}
\usepackage{setspace}

\makeatletter
\AtBeginDocument{%
	\renewcommand*{\AC@hyperlink}[2]{%
		\begingroup
		\hypersetup{hidelinks}%
		\hyperlink{#1}{#2}%
		\endgroup
	}%
}
\makeatother

\makeatletter

\if@twoside 
    \newlength{\textblockoffset}
\fi
\makeatother

\renewcommand{\vec}[1]{\boldsymbol{#1}}

\newcommand{\setR}{\mathbb{R}}

\newcommand{\setZ}{\mathbb{Z}}
\newcommand{\setQ}{\mathbb{Q}}

\newcommand{\ie}{i.e.\ }

\newcommand{\wrt}{w.r.t.\ }

\renewcommand{\d}{\mathrm{d}}

\newcommand{\diag}{\mathrm{diag}}

\newcommand{\idmatrix}{\mathds{1}}

\newcommand{\abs}[1]{\left| #1 \right|}

\newcommand{\teq}{\triangleq}

\let\Vec\relax
\DeclareMathOperator{\Vec}{vec}
\renewcommand{\O}{\mathcal{O}}
\newcommand{\divides}{\mid}
\newcommand{\notdivides}{\nmid}

\newcommand{\GCD}{\ac{GCD}}

\newcommand{\Gps}{\acp{GP}}

\newcommand{\ANNs}{\acp{ANN}}

\newcommand{\fhat}{\widehat{f}}

\newcommand{\alphahat}{\widehat{\alpha}}

\newcommand{\fbar}{\overline{f}}

\newcommand{\setbar}{\mid}

\let\textcite\cite

\DeclarePairedDelimiter\ceil{\lceil}{\rceil}
\DeclarePairedDelimiter\floor{\lfloor}{\rfloor}

\begin{document}

\newacro{ANN}{artificial neural network}
\newacro{CUDA}{Compute Unified Device Architecture}
\newacro{GCD}{greatest common divisor}
\newacro{GP}{Gaussian process}
\newacro{GPN}{Gaussian process neuron}
\newacro{GPU}{graphics processing unit}
\newacro{CDF}{cumulative density function}
\newacro{PDF}{probability density function}
\newacro{MCMC}{Markov chain Monte Carlo}
\newacro{iid}{independent and identical distributed}
\newacro{HMC}{Hamiltonian Monte Carlo}
\newacro{SE}{squared exponential}
\newacro{CNN}{convolutional neural network}
\newacro{RNN}{recurrent neural network}
\newacro{ELBO}{evidence lower bound}
\newacro{DFT}{discrete Fourier transform}

\author{Sebastian Urban\footnote{Technical University Munich, surban@tum.de}, Patrick van der Smagt\footnote{Volkswagen Group, smagt@brml.org}}
\title{Automatic Differentiation for \\ Tensor Algebras \\ {\small Technical Report}}
\date{1 November 2017}

\maketitle
\tableofcontents
\newpage

\section{Introduction}

\textcite{Kjolstad:2017:TAC:3152284.3133901} proposed a tensor algebra compiler.  
It takes expressions that define a tensor element-wise, such as
\[ f_{ij}(a,b,c,\vec{d}) = \exp\left[- \sum_{k=0}^4 \left( (a_{ik} + b_{jk})^2 \, c_{ii} + d_{i+k}^3 \right)  \right] \,, \]
and generates the corresponding compute kernel code.
The arguments can be either dense or sparse matrices.

For machine learning applications, especially deep learning, it is often necessary to compute the gradient of a loss function $l(a,b,c,\vec{d}) = l(f(a,b,c,\vec{d}))$ with respect to model parameters $a, b, c, \vec{d}$.
Hence, if tensor compilers are to be applied in this field, it is necessary to derive expressions for the derivatives of  element-wise defined tensors, \ie expressions of the form $(\d a)_{ik} \teq \partial l / \partial a_{ik}$.

When the mapping between function indices and argument indices is not 1:1, special attention is required.
For example, for the function $f_{ij} (x) = x_i^2$, the derivative of the loss \wrt $x$ is $(\d x)_i \teq \partial l / \partial x_i = \sum_j (\d f)_{ij} \, 2 \, x_i$; the sum is necessary because index $j$ does not appear in the indices of $f$.
Another example is $f_{i} (x) = x_{ii}^2$, where $x$ is a matrix; here we have $(\d x)_{ij} = \delta_{ij} \, (\d f)_i \, 2 \, x_{ii}$; the Kronecker delta is necessary because the derivative is zero for off-diagonal elements.
Another indexing scheme is used by $f_{ij} (x) = \exp x_{i + j}$; here the correct derivative is $(\d x)_{k} = \sum_i (\d f)_{i,k-i} \, \exp x_{k}$, where the range of the sum must be chosen appropriately.

In this publication we present an algorithm that can handle any case in which the indices of an argument are an \emph{arbitrary linear combination} of the indices of the function, thus all of the above examples can be handled.
Sums (and their ranges) and Kronecker deltas are automatically inserted into the derivatives as necessary.
Additionally, the indices are transformed, if required (as in the last example).
The algorithm outputs a symbolic expression that can be subsequently fed into a tensor algebra compiler.

We first review the basic automatic differentiation algorithm (\cref{sec:impl_autodiff_graph,sec:multidim}) and necessary algorithms for integer matrix inversion and for solving systems of linear inequalities (\cref{sec:ieq}).
Then, in \cref{sec:ediff}, we show how to extend automatic differentiation to generate derivative expressions for element-wise defined tensor-valued functions.
An example and numeric verification of our algorithm are presented in \cref{sec:example}.

An open source implementation of the described algorithm is provided at 
\begin{center}
\url{https://github.com/surban/TensorAlgDiff}.
\end{center}
Please cite this publication when using the provided code.

\newpage
\section{Symbolic Reverse Accumulation Automatic Differentiation}\label{sec:impl_autodiff_graph}
\begin{figure}[bp]
\centering
\resizebox{0.7\linewidth}{!}{\begin{tikzpicture}
    \tikzstyle{op}=[circle,blue,very thick,draw,minimum size=17pt,inner sep=0pt,text=black]
    \tikzstyle{deriv}=[rectangle,red,very thick,draw,minimum size=17pt,inner sep=0pt,text=black,fill]
    \tikzstyle{diff}=[circle,red,very thick,draw,minimum size=17pt,inner sep=0pt,text=black]
	\tikzstyle{annot} = [text width=4em, text centered]	
	\tikzstyle{p} = [thick,->,blue]
	\tikzstyle{pd} = [thick,->,red,rounded corners=3pt]

	\def\r{-1.0} \def\c{-0.7} \def\dr{-4.0} \def\drr{4.0}
	\def\dd{\mathrm{d}}

	\node[op,fill,rectangle,text=white] (f) at  (0*\c,0*\r) {$\mathbf{f}$};
	\node[op] (sin1) at  (0*\c,1*\r) {$f_1$};
	\node[op] (sin2) at  (0*\c,2*\r) {$f_2$};
	\node[op] (plus) at  (0*\c,3*\r) {$f_3$};
	\node[op] (mul) at   (1*\c,4*\r) {$f_4$};
	\node[op] (sinh) at  (-1*\c,4*\r) {$f_5$};
	\node[op] (plus2) at (0*\c,5*\r) {$x$};
	
	\path[p] (f) edge (sin1);
	\path[p] (sin1) edge (sin2);
	\path[p] (sin2) edge (plus);
	\path[p] (plus) edge (mul);
	\path[p] (plus) edge (sinh);
	\path[p] (mul) edge (plus2);
	\path[p] (sinh) edge (plus2);	
	
	\node[diff]  (one)  at     (4*\c,1*\r) {$1$};  
	\node[deriv] (dfdsin1) at  (\dr,1*\r) {$\frac{\partial f}{\partial f_1}$};
	\path[pd]  (dfdsin1) edge (one);

	\node[diff]  (dcos1)  at   (2.5*\c,2*\r) {$\frac{\dd f_1}{\dd f_2}$};  
	\node[diff]  (dmul1)  at   (4*\c,2*\r) {$\cdot$};  
	\node[deriv] (dfdsin2) at  (\dr,2*\r) {$\frac{\partial f}{\partial f_2}$};
	\path[pd] (dfdsin2) edge (dmul1);
	\path[pd] (dmul1) edge (dfdsin1);
	\path[pd] (dmul1) edge (dcos1);
	\path[pd] (dcos1) edge (sin2);

	\node[diff]  (dcos2)  at   (2.5*\c,3*\r) {$\frac{\dd f_2}{\dd f_3}$};  
	\node[diff]  (dmul2)  at   (4*\c,3*\r) {$\cdot$};  
	\node[deriv] (dfdplus) at  (\dr,3*\r) {$\frac{\partial f}{\partial f_3}$};
	\path[pd] (dfdplus) edge (dmul2);
	\path[pd] (dmul2) edge (dfdsin2);
	\path[pd] (dmul2) edge (dcos2);
	\path[pd] (dcos2) edge (plus);

	\node[diff]  (dmul3)  at   (4*\c,4*\r) {$\cdot$};  
	\node[diff]  (done)   at   (2.5*\c,4*\r) {$\frac{\partial f_3}{\partial f_4}$};  
	\node[deriv] (dfdmul) at   (\dr,4*\r) {$\frac{\partial f}{\partial f_4}$};
	\path[pd] (dfdmul) edge (dmul3);
	\path[pd] (dmul3) edge (dfdplus);
	\path[pd] (dmul3) edge (done);
	\draw[pd] (done) -- (mul);
	\draw[pd] (done.north) |- ++(-2.2*\c,0.1) -- (sinh.west);
	
	\node[diff]  (dmul4)  at   (4*\c,5*\r) {$\cdot$};  
	\node[diff]  (df4df6) at   (2.5*\c,5*\r) {$\frac{\dd f_4}{\dd x}$};  
	\node[deriv] (dfdc)   at   (0*\c,7*\r) {$\frac{\partial f}{\partial x}$};
	\node[diff]  (dplus6)  at  (0*\c,6*\r) {$+$};  
	\path[pd] (dfdc) edge (dplus6);
	\path[pd] (dmul4) edge (dfdmul);
	\draw[pd] (dmul4) edge (df4df6);
	\draw[pd] (df4df6) edge (plus2);

	\node[diff]  (dmul5)  at   (-4*\c,4*\r) {$\cdot$};  
	\node[diff]  (done2)  at   (-2.5*\c,4*\r) {$\frac{\partial f_3}{\partial f_5}$};  
	\node[deriv] (dfdsinh) at  (\drr,4*\r) {$\frac{\partial f}{\partial f_5}$};
	\path[pd] (dfdsinh) edge (dmul5);
	\path[pd] (dmul5) edge (done2);
	\draw[pd,rounded corners=3pt] (dmul5.north) |- (0,0.6) -| (\dr-0.7,3*\r) -- (dfdplus.west);
	\draw[pd] (done2) -- (sinh);
	\draw[pd] (done2.north) |- ++(2.2*\c,0.1) -- (mul.east);

	\node[diff]  (dmul6)   at   (-4*\c,5*\r) {$\cdot$};  
	\node[diff]  (dcosh)   at   (-2.5*\c,5*\r) {$\frac{\dd f_5}{\dd x}$};  
	\path[pd] (dmul6) edge (dcosh);
	\path[pd] (dmul6) edge (dfdsinh);
	\path[pd] (dcosh) edge (plus2);
	\draw[pd] (dplus6) -| (dmul6);
	\draw[pd] (dplus6) -| (dmul4);

\end{tikzpicture}}
\caption{
The blue nodes show a computational graph for the function $f(x)=f_1 \big[f_2 \big(f_3 \big( f_4(x), f_5(x) \big) \big) \big]$. 
Each node $f_1$, $f_2$, $\dots$, $f_5$ represents a part of the function and each edge represents an argument.
By applying automatic differentiation as described in \cref{sec:impl_autodiff_graph} the computational graph for the derivatives (shown in red) is obtained.
}
\label{fig:impl_autodiff}
\end{figure}
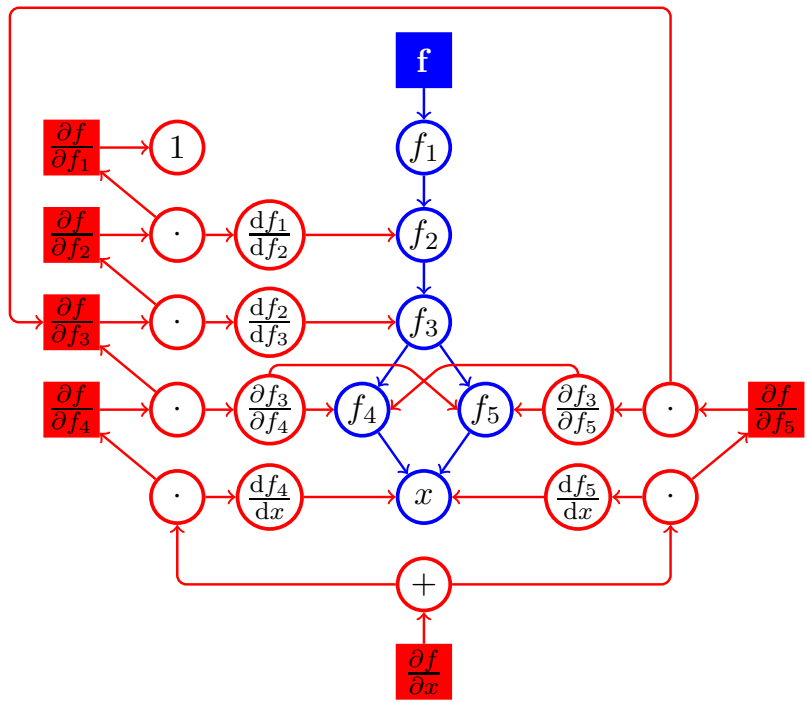

Every function $f$ can be written as a composition of elementary functions such as addition, subtraction, multiplication, division, trigonometric function, the exponential function, the logarithm and so on.
For now let us assume that the elementary functions take one or more \emph{scalar} arguments; thus $f$ will also be a function accepting scalar arguments.
For example, the function $f(x_1,x_2) = \exp(x_1 + x_2)$ can be written as $f(x_1,x_2) = f_1(f_2(x_1,x_2))$ with parts $f_1(t) = \exp(t)$ and $f_2(t_1,t_2) = t_1+t_2$.
It is also possible that parts appear more than once in a function.
As an example $f(x) = \sin(x^2) \cdot \cos(x^2)$ can be decomposed into $f(x) = f_1\big[f_2\big( f_4(x) \big), f_3\big(f_4(x)\big) \big]$ where the parts $f_1(s, t) = s \cdot t$, $f_2(t) = \sin(t)$, $f_3(t) = cos(t)$ are used once and $f_4(t) = t^2$ is used twice. 
A decomposition of a function into parts can be represented by a computational graph, that is a directed acyclic graph where each node represents a function part $f_i$ and an edge between two node represents that the target node is used as the value to an argument of the source node.
An exemplary computational graph for the function $f(x)=f_1 \big[f_2 \big(f_3 \big( f_4(x), f_5(x) \big) \big) \big]$ is shown by the blue nodes in \cref{fig:impl_autodiff}.

Automatic differentiation is based on the well-known chain rule, which states that for a scalar function of the form $f(x) = g(h(x))$ the derivative can be written as
\[ \frac{\d f}{\d x} =  \frac{\partial g}{\partial h} \, \frac{\partial h}{\partial x} \,. \]
Given a function $f$ and its decomposition into parts $f_i$, the following algorithm uses reverse accumulation automatic differentiation to obtain a computational graph for the derivatives of $f$.
Since $f(\vec{x}) = f_1(\dots)$ the derivative of $f$ \wrt $f_1$ is
\begin{equation}\label{eq:impl_autodiff_ff1}
\frac{\partial f}{\partial f_1} = 1 \,. 
\end{equation}
Then iteratively do the following:
Find a part $f_i$ for which the derivative of all its consumers is available but $\partial f / \partial f_i$ is yet unknown.
A part $f_c$ is a consumer of part $f_i$, if $f_i$ occurs as a direct argument to $f_c$ in the function $f$.
Thus, in the graphical representation of $f$ part $f_c$ is a consumer of $f_i$, if there exists an edge from $f_c$ to $f_i$.
Since the computational graph of a function is acyclic, there will always exist a part $f_i$ for which this condition is fulfilled.
Let $\mathrm{csmr}(f_i)$ be the set of consumers of part $f_i$.
Following the chain rule, the derivative of $f$ \wrt $f_i$ is given by
\begin{equation}\label{eq:impl_autodiff_ffi}
\frac{\partial f}{\partial f_i} = \sum_{d \in \mathrm{csmr}(f_i)} \frac{\partial f}{\partial f_d} \frac{\partial f_d}{\partial f_i} \,.
\end{equation}
Repeat this process until the derivatives \wrt all parts $\partial f / \partial f_i$ have been calculated.
Once completed, the derivatives of $f$ \wrt its arguments $x_j$, $j \in \{1, \dots, n\}$, follow immediately,
\begin{equation}\label{eq:impl_autodiff_fxj}
\frac{\partial f}{\partial x_j} = \sum_{d \in \mathrm{csmr}(x_j)} \frac{\partial f}{\partial f_d} \frac{\partial f_d}{\partial x_j} \,.
\end{equation}
Note, that this algorithm requires a single pass only to complete the derivatives of $f$ \wrt to \emph{all} of its parameters.

By performing this algorithm on the computational graph shown in \cref{fig:impl_autodiff}, the derivative represented by the red nodes and edges is obtained.
The computation proceeds from top to bottom in a breadth-first order of traversation.
In general the partial derivatives of the function parts can depend on all of its arguments, as it can be seen in the dependencies of the nodes for $\partial f_3 / \partial f_4$ and $\partial f_3 / \partial f_5$.
Symbolic derivatives can be obtained from the resulting computational graph by starting from the node $\partial f / \partial x_i$ and following the dependencies until reaching the leafs of the graph.
However, for numerical evaluation it is more efficient to insert numerical values for the parameters $\vec{x}$ into the graph and then evaluate it node by node.
This ensures that intermediate values are only computed once and thus the possibility of an exponential blow up of the number of terms that can occur during classical symbolic differentiation is avoided.
To evaluate the derivative $\partial f / \partial x$ numerically, the function $f(x)$ must be evaluated followed by the derivatives of all parts.
This corresponds to the forward and backward passes of the backpropagation algorithm for neural networks.

\begin{figure}[tb]
\centering
\resizebox{0.8\linewidth}{!}{\begin{tikzpicture}
    \tikzstyle{op}=[circle,blue,very thick,draw,minimum size=17pt,inner sep=0pt,text=black]
    \tikzstyle{deriv}=[rectangle,red,very thick,draw,minimum size=17pt,inner sep=0pt,text=black,fill]
    \tikzstyle{diff}=[circle,red,very thick,draw,minimum size=17pt,inner sep=0pt,text=black]
	\tikzstyle{annot} = [text width=4em, text centered]	
	\tikzstyle{p} = [thick,->,blue]
	\tikzstyle{pd} = [thick,->,red]

	\def\r{-1.0} \def\c{-0.7} \def\dr{-4.0} \def\drr{4.8}

	\node[op,fill,rectangle,text=white] (f) at  (0*\c,0*\r) {$\mathbf{f}$};
	\node[op] (sin1) at  (0*\c,1*\r) {\footnotesize $\sin_1$};
	\node[op] (sin2) at  (0*\c,2*\r) {\footnotesize $\sin_2$};
	\node[op] (plus) at  (0*\c,3*\r) {$+_1$};
	\node[op] (mul) at   (1*\c,4*\r) {$\cdot$};
	\node[op] (sinh) at  (-1*\c,4*\r) {\footnotesize $\sinh$};
	\node[op] (c) at     (2*\c,5*\r) {$x_1$};
	\node[op] (plus2) at (0*\c,5*\r) {$+_2$};
	\node[op] (a) at     (1*\c,7*\r) {$x_2$};
	\node[op] (b) at     (-1*\c,7*\r) {$x_3$};
	
	\path[p] (f) edge (sin1);
	\path[p] (sin1) edge (sin2);
	\path[p] (sin2) edge (plus);
	\path[p] (plus) edge (mul);
	\path[p] (plus) edge (sinh);
	\path[p] (mul) edge (c);	
	\path[p] (mul) edge (plus2);
	\path[p] (sinh) edge (plus2);	
	\path[p] (plus2) edge (a);	
	\path[p] (plus2) edge (b);	
	
	\node[diff]  (one)  at     (4*\c,1*\r) {$1$};  
	\node[deriv] (dfdsin1) at  (\dr,1*\r) {$\frac{\partial f}{\partial\sin_1}$};
	\path[pd]  (dfdsin1) edge (one);

	\node[diff]  (dcos1)  at   (2.5*\c,2*\r) {\footnotesize $\cos$};  
	\node[diff]  (dmul1)  at   (4*\c,2*\r) {$\cdot$};  
	\node[deriv] (dfdsin2) at  (\dr,2*\r) {$\frac{\partial f}{\partial\sin_2}$};
	\path[pd] (dfdsin2) edge (dmul1);
	\path[pd] (dmul1) edge (dfdsin1);
	\path[pd] (dmul1) edge (dcos1);
	\path[pd] (dcos1) edge (sin2);

	\node[diff]  (dcos2)  at   (2.5*\c,3*\r) {\footnotesize $\cos$};  
	\node[diff]  (dmul2)  at   (4*\c,3*\r) {$\cdot$};  
	\node[deriv] (dfdplus) at  (\dr,3*\r) {$\frac{\partial f}{\partial +_1}$};
	\path[pd] (dfdplus) edge (dmul2);
	\path[pd] (dmul2) edge (dfdsin2);
	\path[pd] (dmul2) edge (dcos2);
	\path[pd] (dcos2) edge (plus);

	\node[diff]  (dmul3)  at   (4*\c,4*\r) {$\cdot$};  
	\node[diff]  (done)   at   (2.5*\c,4*\r) {$1$};  
	\node[deriv] (dfdmul) at   (\dr,4*\r) {$\frac{\partial f}{\partial\,\cdot}$};
	\path[pd] (dfdmul) edge (dmul3);
	\path[pd] (dmul3) edge (dfdplus);
	\path[pd] (dmul3) edge (done);
	
	\node[diff]  (dmul4)  at   (4*\c,5*\r) {$\cdot$};  
	\node[deriv] (dfdc)   at   (\dr,5*\r) {$\frac{\partial f}{\partial x_1}$};
	\path[pd] (dfdc) edge (dmul4);
	\path[pd] (dmul4) edge (dfdmul);
	\draw[pd,rounded corners=3pt] (dmul4.east) -| (2.6*\c,5.7*\r) -- (1.4*\c,5.7*\r) |- (plus2);

	\node[diff]  (dmul5)  at   (-4*\c,4*\r) {$\cdot$};  
	\node[diff]  (done2)  at   (-2.5*\c,4*\r) {$1$};  
	\node[deriv] (dfdsinh) at  (\drr,4*\r) {$\frac{\partial f}{\partial\sinh}$};
	\path[pd] (dfdsinh) edge (dmul5);
	\path[pd] (dmul5) edge (done2);
	\draw[pd,rounded corners=3pt] (dmul5.north) |- (0,0.6) -| (\dr-0.7,3*\r) -- (dfdplus.west);

	\node[diff]  (dplus6)  at  (-5.5*\c,5.5*\r) {$+$};  
	\node[diff]  (dmul6)   at   (-4*\c,5*\r) {$\cdot$};  
	\node[diff]  (dcosh)   at   (-2.5*\c,5*\r) {\footnotesize $\cosh$};  
	\node[diff]  (dmul7)   at   (-4*\c,6*\r) {$\cdot$};  
	\node[deriv] (dfdplus2) at  (\drr,5.5*\r) {$\frac{\partial f}{\partial +_2}$};
	\path[pd] (dfdplus2) edge (dplus6);
	\path[pd] (dplus6) edge (dmul6);
	\path[pd] (dplus6) edge (dmul7);
	\path[pd] (dmul6) edge (dcosh);
	\path[pd] (dmul6) edge (dfdsinh);
	\path[pd] (dcosh) edge (plus2);
	\draw[pd,rounded corners=3pt] (dmul7.west) -| (-0.9*\c+0.5,7*\r-0.5) -| (1.8*\c,6.9*\r) |- (2.25*\c,5.8*\r) |- (2.0*\c,5.6*\r) -- (c.south);
	\draw[pd,rounded corners=3pt] (dmul7.south west) -| (-1.15*\c+0.5,7*\r-0.7) -| (\dr-0.7,4*\r) -- (dfdmul.west);

	\node[diff]  (dmul8)   at   (-4*\c,7*\r) {$\cdot$};  
	\node[diff]  (done5)  at   (-2.5*\c,7*\r) {$1$};  
	\node[deriv] (dfdb)    at  (\drr,7*\r) {$\frac{\partial f}{\partial x_3}$};
	\path[pd] (dfdb) edge (dmul8);
	\path[pd] (dmul8) edge (done5);
	\path[pd] (dmul8) edge (dfdplus2);

	\node[diff]  (dmul9)  at   (4*\c,7*\r) {$\cdot$};  
	\node[diff]  (done4)  at   (2.5*\c,7*\r) {$1$};  
	\node[deriv] (dfda)   at   (\dr,7*\r) {$\frac{\partial f}{\partial x_2}$};
	\path[pd] (dfda) edge (dmul9);
	\path[pd] (dmul9) edge (done4);
	\draw[pd,rounded corners=3pt] (dmul9.south) |- (4.3*\c,7*\r-0.6) |- (4.3*\c,7*\r-0.8) -| (4*\c,7*\r-1.2) -| (\drr+0.7,7*\r-1.2) |- (dfdplus2.east);

\end{tikzpicture}}
\caption{
A practical example for automatic symbolic differentiation.
The computational graph for the function $f(x_1, x_2, x_3) = \sin\!\big[ \sin\!\big( x_1 \cdot (x_2 + x_3) + \sinh(x_2 + x_3) \big) \big]$ is shown in blue.
The computational graph for the derivatives obtained by automatic differentiation is shown in red.
Note how intermediate values are reused automatically and the derivatives \wrt different $x_i$ share most parts of the computational graph.
Symbolic derivatives can be extracted from the graph or it can be evaluated numerically by substituting values for $x_1$, $x_2$ and $x_3$.
}
\label{fig:impl_autodiff_example}
\end{figure}
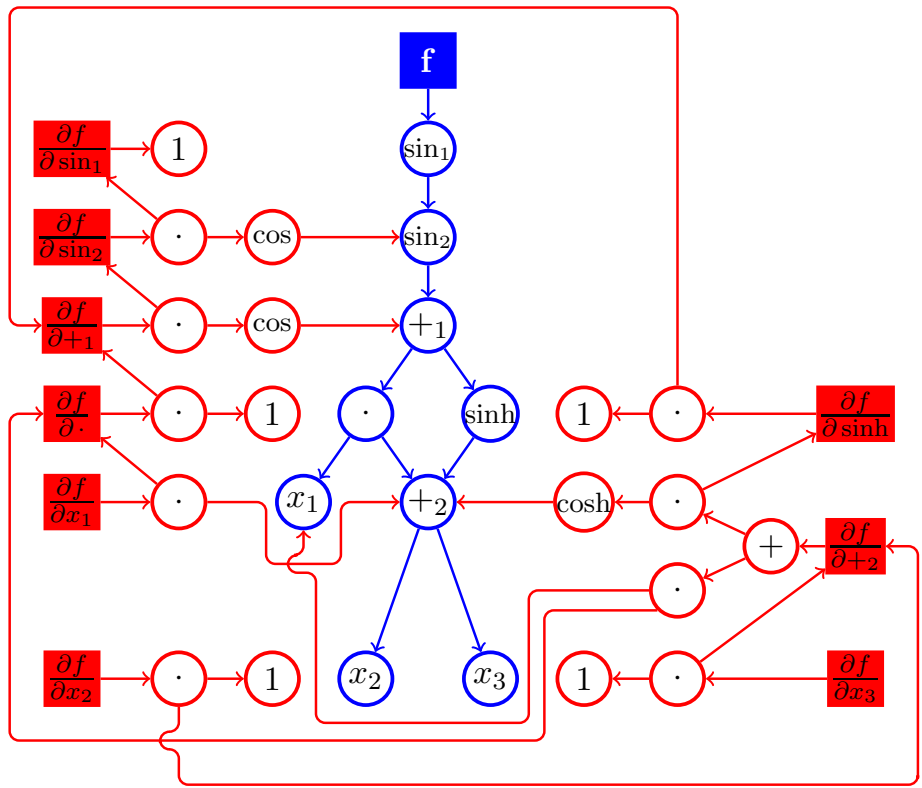

An example of a computational graph and its derivative for the concrete function
\[ f(x_1, x_2, x_3) = \sin\!\big[ \sin\!\big( x_1 \cdot (x_2 + x_3) + \sinh(x_2 + x_3) \big) \big] \]
is shown in \cref{fig:impl_autodiff_example}.

\clearpage
\section{Handling Multidimensional Functions}\label{sec:multidim}
So far we have shown automatic differentiation for scalar functions.
However, in the context of \ANNs{} and \Gps{} we will mostly be dealing with functions that deal with tensor-valued functions.
While any tensor-valued function can be written as a scalar function by splitting it into separate functions for each element of the tensor, doing so often has a significant penalty on computational efficiency.
For example consider matrix multiplication.
Calculating each element of $C = A \cdot B$ separately using $C_{ij} = \sum_k A_{ik}\,B_{kj}$ requires a total of $\O(n^3)$ operations where $n$ is the size of the square matrices $A$ and $B$.
Contrary to that calculating all elements simultaneously can be done in $\O(n^{2.807})$ using the Strassen algorithm \cite{strassen1969gaussian} or even more efficiently in $\O(n^{2.375})$ using the Coppersmith-Winograd algorithm \cite{coppersmith1987matrix}.\footnote{While these algorithms are asymptotically faster than naive matrix multiplication, they also have a larger constant factor in their running time not captured by the big O notation. Therefore in practice they are only beneficial for matrices larger than a certain size.}
Thus we will show how to perform automatic differentiation on multidimensional functions now.

For functions working in two- or higher dimensional space, we use the vectorization operator $\Vec$ to transform them into vector-valued functions.
For a $D$-dimensional tensor $A \in \setR^{N_1 \times N_2 \times \cdots \times N_D}$ the vectorization operator is defined element-wise by
\begin{equation}\label{eq:impl_vectorization}
(\Vec A)_{\sum_d s_d i_d} = A_{i_1, i_2, \dots, i_D} \,, \quad\quad i_d \in \{1, \dots, N_d\} 
\end{equation}
where the strides $\vec{s}$ are given by
\[ s_d = \prod_{b=2}^d N_{b-1} \,. \]
As an example, for a matrix $A \in \setR^{N \times M}$ this operator takes the columns of the matrix and stacks them on top of one another,
\[ \Vec A = \big( A_{11}, A_{21}, \dots, A_{N1}, A_{12}, A_{22}, \dots, A_{N2}, \dots, A_{1M}, A_{2M}, \dots, A_{NM} \big)^T \,. \]
Thus the derivatives of a tensor-valued function $F: \setR^{N_1 \times N_2 \times \cdots \times N_D} \to \setR^{M_1 \times M_2 \times \cdots \times M_{D'}}$ can be dealt with by defining a helper function $\widehat{F}: \setR^{N_1 \, N_2 \cdots N_D} \to \setR^{M_1 \, M_2 \cdots M_{D'}}$ with $\widehat{F}(\Vec X) = \Vec F(X)$ and considering the derivatives of this vector-valued function $\widehat{F}$ instead.

It remains to show how to apply automatic differentiation to vector-valued functions.
To do so, let us us first see how the chain rule works on vector-valued functions.
Consider two functions, $\vec{g}: \setR^K \to \setR^N$ and $\vec{h}: \setR^M \to \setR^K$, and a composite function $\vec{f}: \setR^M \to \setR^N$ with $\vec{f}(\vec{x}) = \vec{g}( \vec{h} ( \vec{x} ) )$.
By expanding $\vec{g}(\vec{r})$ as $\vec{g}(r_1, r_2, \dots, r_K)$ and $\vec{h}(\vec{x})$ as $\big( h_1(\vec{x}), h_2(\vec{x}), \dots, h_K(\vec{x}) \big)^T$ we can write 
\[ f_i(\vec{x}) = g_i\big( h_1(\vec{x}), h_2(\vec{x}), \dots, h_K(\vec{x}) \big) \]
and apply the chain rule on each argument of $g_i$, resulting in
\begin{equation}\label{eq:autodiff_cr_vec_elems}
\frac{\partial f_i}{\partial x_j} = \sum_{k=1}^K \frac{\partial g_i}{\partial h_k} \frac{\partial h_k}{\partial x_j} \,.
\end{equation}
By introducing the Jacobian
\[ \left( \frac{\d \vec{f}}{\d \vec{x}} \right)_{ij}  \teq \frac{\partial f_i}{\partial x_j} \]
we can rewrite \eqref{eq:autodiff_cr_vec_elems} as a vectorized equation,
\begin{equation}\label{eq:autodiff_cr_vec}
\frac{\d \vec{f}}{\d \vec{x}} = \frac{\partial \vec{g}}{\partial \vec{h}} \frac{\partial \vec{h}}{\partial \vec{x}} \,,
\end{equation}
and thus obtain the chain rule for vector-valued functions.
As we see, it is like the chain rule for scalars but with scalar multiplication replaced by matrix multiplication.

The algorithm for automatic differentiation for vector-valued functions is thus equal to scalar automatic differentiation described in \cref{sec:impl_autodiff_graph}, but with \cref{eq:impl_autodiff_ff1} replaced by
\begin{equation}
\frac{\partial \vec{f}}{\partial \vec{f_1}} = \idmatrix
\end{equation}
and \cref{eq:impl_autodiff_ffi} replaced by
\begin{equation}\label{eq:autodiff_prop_vec}
\frac{\partial \vec{f}}{\partial \vec{f_i}} = \sum_{d \in \mathrm{csmr}(\vec{f_i})} \frac{\partial \vec{f}}{\partial \vec{f_d}} \frac{\partial \vec{f_d}}{\partial \vec{f_i}} \,.
\end{equation}
For many common operations the size of the Jacobian $\partial \vec{f_d} / \partial \vec{f_i}$ may become very large.
For example, the Jacobian of a matrix multiplication is of size $n^4$ for two matrices of size $n \times n$.
However, since most elements are indeed zero, it is possible and vastly more efficient to directly compute the product $(\partial \vec{f} / \partial \vec{f_d}) (\partial \vec{f_d} / \partial \vec{f_i})$ without explicitly evaluating the Jacobian.
This is also the case for all elementary operations that work element-wise, such as addition, subtraction and the Hadamard product, which result in a diagonal Jacobian matrix.
Consequently the explicit form \eqref{eq:autodiff_prop_vec} should only be used as a fall-back when such a shortcut computation is not available.

\section{Systems of Integer Equalities and Inequalities}\label{sec:ieq}
This section introduces methods to solve systems of integer equalities and inequalities.
The algorithms presented here will be employed the compute the element-wise derivative expressions of tensor-valued functions.

\subsection{Systems of Linear Integer Equations}\label{sec:integer_eqs}
Consider a system of linear equations 
\begin{align*}
A_{11} \, x_1 + A_{12} \, x_2 + \cdots + A_{1m} \, x_m &= b_1 \\
A_{21} \, x_1 + A_{22} \, x_2 + \cdots + A_{2m} \, x_m &= b_2 \\
\vdots                                                 &\vdots \vdots \\
A_{n1} \, x_1 + A_{n2} \, x_2 + \cdots + A_{nm} \, x_m &= b_n  \,,
\end{align*}
with integer coefficients $A \in \setZ^{N \times M}$, integer variables $\vec{x} \in \setZ^M$ and integer targets $\vec{b} \in \setZ^N$.
In matrix notation this system can be expressed much briefer as
\begin{equation}\label{eq:integer_system}
A \, \vec{x} = \vec{b} \,.
\end{equation}
To determine the set of solutions the matrix $A$ must be transformed into Smith normal form, which is a diagonal matrix of the form
\begin{equation}\label{eq:smith_nf}
S = \diag(\alpha_1, \alpha_2, \dots, \alpha_R, 0, \dots, 0)
\end{equation}
with the property that 
\begin{equation}\label{eq:smith_nf_div}
\alpha_i \divides \alpha_{i+1}, \quad 1 \leq i < r \,,
\end{equation}
where $a \divides b$ should be read as ``$a$ \emph{divides} $b$''.
Analogously $a \notdivides b$ should be read as ``$a$ \emph{does not divide} $b$''.
The number of non-zero entries $R$ in the diagonal corresponds to the rank of $A$.
It can be shown \cite{adkins} that for each non-zero matrix $A \in \setZ^{N \times M}$ there exist invertible matrices $U \in \setZ^{N\times N}$ and $V \in \setZ^{M\times M}$ so that
\begin{equation}\label{eq:smith_trafo}
S = U \, A \, V
\end{equation}
where $S \in \setZ^{N \times M}$ is the Smith normal form of $A$.
Using the smith normal form, the equation system \eqref{eq:integer_system} can be rewritten as
\begin{equation}\label{eq:smith_eqs}
S \, \vec{x}' = \vec{b}'
\end{equation}
with
\begin{align}
\vec{x}  &= V \, \vec{x}' \,, \\
\vec{b}' &= U \, \vec{b} \,.
\end{align}
Since $S$ is diagonal, the solutions can be read off from \eqref{eq:smith_eqs}, as we describe in the following.

For the zero rows of $S$ the corresponding entries of $\vec{b}'$ must also be zero, otherwise the equation system would be inconsistent and no solution exists.
Thus for the system to be solvable we must have
\begin{equation}
C \, \vec{b} = \vec{0}
\end{equation}
where $C \in \setZ^{N-R \times N}$ with $C_{ij} = U_{R+i,j}$ is the sub-matrix consisting of the rows $R+1$ to $N$ of $U$.
It is called the cokernel of $A$.

For each non-zero entry $\alpha_i$ of $S$ we must have 
\begin{equation}
x'_i = \frac{b'_i}{\alpha_i}
\end{equation}
and thus a solution exists only if $b'_i$ is dividable by $\alpha_i$.
We can define a so-called pseudo-inverse $I: \setQ^{M \times N}$ with
\begin{equation}
I \teq V \, S^{\dagger} \, U
\end{equation} 
where $S^{\dagger} \in \setQ^{N \times M}$ is defined by
\begin{equation}
S^{\dagger} = \diag(1/\alpha_1, 1/\alpha_2, \dots, 1/\alpha_R, 0, \dots, 0) \,,
\end{equation}
with the factors $\alpha_i$ given by \eqref{eq:smith_nf}.
This pseudo-inverse has the property that $A \, I \, A = A$.
Thus, for every $\vec{b}$ that is in the cokernel of $A$, we can obtain an $\vec{x}$ by setting $\vec{x} = I \, \vec{b}$ so that $A \, \vec{x} = \vec{b}$.

For the zero columns of S the corresponding entries of $\vec{x}'$ do not affect the value of $\vec{b}'$.
Consequently, the columns of the matrix $K \in \setZ^{M \times M-R}$, with $K_{ij} = V_{i,R+j}$, are a basis for the kernel (also called null-space) of $A$.
This means that $M \, K = \vec{0}$ and thus every $\vec{b}$ that is in the cokernel of $A$ we can write $\vec{b} = A( I \, \vec{b} + K \, \vec{z} )$ where $\vec{z} \in \setZ^{M-R}$ is a vector of arbitrary integers.

In summary, the equation system $A \, \vec{x} = \vec{b}$ has no integer solution for a particular $\vec{b}$, if $C \, \vec{b} \neq \vec{0}$ or $I \, \vec{b} \notin \setZ^N$.
Otherwise, if $A$ has full rank, that is $R=N=M$, a unique integer solution exists, determined by $\vec{x} = I \, \vec{b}$.
If $A$ has non-full rank, infinitely many integer solutions exist and are given by $\vec{x} = I \, \vec{b} + K \, \vec{z}$ where $\vec{z} \in \setZ^{M-R}$ is a vector of arbitrary integers.

\subsubsection{Computation of the Smith Normal Form}
An algorithm \cite{smith1860systems} that, given a matrix $A$, computes the Smith normal form $S$ and two matrices $U$ and $V$, such that $S = U \, A \, V$ is shown in \cref{alg:smith}.
The algorithm transforms the matrix $A$ into Smith normal form by a series of elementary row and column operations.
Matrices $U$ and $V$ are initialized to be identity matrices and the same row and column operations are applied to them, so that in the end the relation $S = U \, A \, V$ holds.
Since all operations are elementary, it follows that $U$ and $V$ are invertible as required.
By following the description of the algorithm it is clear that the resulting matrix $S$ will be diagonal and fulfill the property \eqref{eq:smith_nf_div}.
To find the factors $\beta$, $\sigma$ and $\tau$ of B\'ezout's identity in steps \ref{alg:smith_euclid1} and \ref{alg:smith_euclid2} the extended Euclidean algorithm \cite{TAOCP2} is used, which is shown in \cref{alg:extended_euclid}.

What remains to show is that the described algorithm terminates.
With each iteration of the loop in step \ref{alg:smith_s_changing} the absolute value of the element $S_{aa}$ decreases, because it is replaced with the \GCD{} of itself and another element.
Thus, this loop will terminate since, in worst case, $S_{aa}=+1$ or $S_{aa}=-1$ will divide all following rows and columns.
The same argument holds, when the matrix must be rediagonalized due to the execution of step \ref{alg:smith_rerun}.
It is easy to verify that the first diagonalization step executed thereafter will set $S_{aa} = \mathrm{gcd}(S_{aa}, S_{a+1,a+1})$ and thus the absolute value of $S_{aa}$ decreases.
Thus, in the worst case, the loop terminates as soon as $S_{11} = S_{22} = \cdots = S_{R-1,R-1} = 1$, which then divides $S_{RR}$.

\begin{algorithm}[tbp]\DontPrintSemicolon
\caption{Smith normal form of an integer matrix}\label{alg:smith}
\KwIn{non-zero matrix $A \in \setZ^{N \times M}$}
\KwOut{Smith normal form $S \in \setZ^{N \times M}$, invertible matrices $U \in \setZ^{N \times N}$, $V \in \setZ^{M \times M}$, rank $R$}
\vspace{2mm}
$U \longleftarrow \idmatrix_N$; $V \longleftarrow \idmatrix_M$ 		\tcp*{initialize $U$ and $V$ with identity matrices} 
$S \longleftarrow A$ 												\tcp*{initialize $S$ with $A$} 
$a \longleftarrow 1$ 												\tcp*{initialize active row and column}
\While(\label{alg:smith_diagonalize}\tcp*[f]{diagonalize $S$}){$\exists i,j: i \geq a \,\land\, j \geq a \,\land\, S_{ij} \neq 0$}{
	\tcp{Bring non-zero pivot element into position $S_{aa}$}
	Simultaneously $S_{\cdot a} \longleftarrow S_{\cdot j}$ and $S_{\cdot j} \longleftarrow S_{\cdot a}$ \\
	Simultaneously $V_{\cdot a} \longleftarrow V_{\cdot j}$ and $V_{\cdot j} \longleftarrow V_{\cdot a}$ \\
	Simultaneously $S_{a \cdot} \longleftarrow S_{i \cdot}$ and $S_{i \cdot} \longleftarrow S_{a \cdot}$ \\	
	Simultaneously $U_{a \cdot} \longleftarrow U_{i \cdot}$ and $U_{i \cdot} \longleftarrow U_{a \cdot}$ \\		
	\While(\label{alg:smith_s_changing}\tcp*[f]{zero all elements below and right of $S_{aa}$}){$S$ is changing}{
		\While(\tcp*[f]{ensure divisibility of rows}){$\exists i: i > a \,\land\, S_{aa} \notdivides S_{ia}$}{
			\label{alg:smith_euclid1} Find $\beta, \sigma, \tau$ so that $\beta = \mathrm{gcd}(S_{aa}, S_{ia}) = \sigma \, S_{aa} + \tau \, S_{ia}$. \\
			$\gamma \longleftarrow \frac{S_{ia}}{\beta}$; $\quad \alpha \longleftarrow \frac{S_{aa}}{\beta}$ \\
			Simultaneously $S_{a\cdot} \longleftarrow \sigma \, S_{a\cdot} + \tau   S_{i\cdot}$ and $S_{i\cdot} \longleftarrow -\gamma \, S_{a\cdot} + \alpha S_{i\cdot}$ \\
			Simultaneously $U_{a\cdot} \longleftarrow \sigma \, U_{a\cdot} + \tau   U_{i\cdot}$ and $U_{i\cdot} \longleftarrow -\gamma \, U_{a\cdot} + \alpha U_{i\cdot}$ 				
		} 
		\While(\tcp*[f]{eliminate first element of rows}){$\exists i: i > a \,\land\, S_{ia} \neq 0$}{ 
			$f \longleftarrow \frac{S_{ia}}{S_{aa}}$ \\
			$S_{i\cdot} \longleftarrow S_{i\cdot} - f \, S_{a\cdot}$ \\
			$U_{i\cdot} \longleftarrow U_{i\cdot} - f \, U_{a\cdot}$ 
		}
		\While(\tcp*[f]{ensure divisibility of columns}){$\exists j: j > a \,\land\, S_{aa} \notdivides S_{aj}$}{
			\label{alg:smith_euclid2} Find $\beta, \sigma, \tau$ so that $\beta = \mathrm{gcd}(S_{aa}, S_{aj}) = \sigma \, S_{aa} + \tau \, S_{aj}$. \\
			$\gamma \longleftarrow \frac{S_{aj}}{\beta}$; $\quad \alpha \longleftarrow \frac{S_{aa}}{\beta}$ \\
			Simultaneously $S_{\cdot a} \longleftarrow \sigma \, S_{\cdot a} + \tau   S_{\cdot j}$ and $S_{\cdot j} \longleftarrow -\gamma \, S_{\cdot a} + \alpha S_{\cdot j}$ \\
			Simultaneously $V_{\cdot a} \longleftarrow \sigma \, V_{\cdot a} + \tau   V_{\cdot j}$ and $V_{\cdot j} \longleftarrow -\gamma \, V_{\cdot a} + \alpha V_{\cdot j}$ 				
		} 
		\While(\tcp*[f]{eliminate first element of columns}){$\exists j: j > a \,\land\, S_{aj} \neq 0$}{ 
			$f \longleftarrow \frac{S_{aj}}{S_{aa}}$ \\
			$S_{\cdot j} \longleftarrow S_{\cdot j} - f \, S_{\cdot a}$ \\
			$V_{\cdot j} \longleftarrow V_{\cdot j} - f \, V_{\cdot a}$ 
		}	
	}
	$a \longleftarrow a+1$ \tcp*{next diagonal element}
}
$R \longleftarrow a-1$ \tcp*{rank is number of non-zero diagonal elements}
\For{$a \in \{1, \dots, R\}$}{
	\If(\tcp*[f]{ensure positive diagonal}){$S_{aa} < 0$}{
		$S_{\cdot a} \longleftarrow -S_{\cdot a}$ \\
		$V_{\cdot a} \longleftarrow -V_{\cdot a}$
	}
	\If(\tcp*[f]{ensure divisibility constraints}){$a \leq R-1 \,\land\, S_{aa} \notdivides S_{a+1,a+1}$}{
		$S_{\cdot a} \longleftarrow S_{\cdot a} + S_{\cdot, a+1}$ \\
		$V_{\cdot a} \longleftarrow V_{\cdot a} + V_{\cdot, a+1}$ \\		
		\label{alg:smith_rerun} Go back to step \ref{alg:smith_diagonalize}.
	}
}
\end{algorithm}

\begin{algorithm}[tbp]\DontPrintSemicolon
\caption{Extended Euclidean algorithm}\label{alg:extended_euclid}
\KwIn{positive numbers $a \in \setZ^+$, $b \in \setZ^+$}
\KwOut{factors $x \in \setZ$, $y \in \setZ$, $z \in \setZ$ fulfilling B\'ezout's identity $z = \mathrm{gcd}(a,b) = a\,x + b\,y$}
\vspace{2mm}
$r_0 \longleftarrow a$ ; $r_1 \longleftarrow b$ \\
$s_0 \longleftarrow 1$ ; $s_1 \longleftarrow 0$ \\
$t_0 \longleftarrow 0$ ; $t_1 \longleftarrow 1$ \\
$i \longleftarrow 1$ \\
\While{$r_i \neq 0$}{
	$q \longleftarrow \frac{r_{i-1}}{r_i}$ \tcp*{integer division}
	$r_{i+1} \longleftarrow r_{i-1} - q \, r_i$ \\
	$s_{i+1} \longleftarrow s_{i-1} - q \, s_i$ \\	
	$t_{i+1} \longleftarrow t_{i-1} - q \, t_i$ \\		
	$i \longleftarrow i+1$
}
$z \longleftarrow r_{i-1}$; $x \longleftarrow s_{i-1}$; $y \longleftarrow t_{i-1}$
\end{algorithm}

\subsection{Systems of Linear Inequalities}\label{sec:lin_ieq}
Consider a system of linear inequalities
\begin{align}
A_{11} \, x_1 + A_{12} \, x_2 + \cdots + A_{1M} \, x_M &\geq b_1 \nonumber \\
A_{21} \, x_1 + A_{22} \, x_2 + \cdots + A_{2M} \, x_M &\geq b_2 \nonumber \\
\vdots                                                 &\vdots \vdots  \label{eq:ieq_orignal} \\
A_{N1} \, x_1 + A_{N2} \, x_2 + \cdots + A_{NM} \, x_M &\geq b_N \nonumber \,,
\end{align}
with coefficients $A \in \setR^{N \times M}$, variables $\vec{x} \in \setR^M$ and biases $\vec{b} \in \setR^N$.
Note that this notation can also describe equalities by including the same line twice, where one occurrence is multiplied by $-1$ on both sides.
In matrix notation this inequality system can be expressed much briefer as
\begin{equation}\label{eq:ieq_system}
A \, \vec{x} \geq \vec{b} \,.
\end{equation}
The objective is to transform the inequality system into the form
\begin{alignat}{2}
\max (L^M \vec{b}) 										                &\leq x_M     &&\leq \min (H^M \vec{b})  \label{eq:ieq_tf_m} \\
\max (L^{M-1} \vec{b} + \widehat{L}^{M-1} \vec{x}_{M})                  &\leq x_{M-1} &&\leq \min (H^{M-1} \vec{b} + \widehat{H}^{M-1} \vec{x}_{M}) \\
\max (L^{M-2} \vec{b} + \widehat{L}^{M-2} \vec{x}_{M-1 \,\dots\, M})    &\leq x_{M-2} &&\leq \min (H^{M-2} \vec{b} + \widehat{H}^{M-2} \vec{x}_{M-1 \,\dots\, M}) \\
\vdots & \vdots && \vdots\vdots \nonumber \\
\max (L^{2} \vec{b} + \widehat{L}^{2} \vec{x}_{3 \,\dots\, M})          &\leq x_{2}   &&\leq \min (H^{2} \vec{b} + \widehat{H}^{2} \vec{x}_{3 \,\dots\, M}) \\
\max (L^{1} \vec{b} + \widehat{L}^{1} \vec{x}_{2 \,\dots\, M})          &\leq x_{1}   &&\leq \min (H^{1} \vec{b} + \widehat{H}^{1} \vec{x}_{2 \,\dots\, M}) \,, \label{eq:ieq_tf_1}
\end{alignat}
so that the range of each element $x_i$ can be determined sequentially.
Here $\vec{x}_{i \,\dots\, j}$ should be read as the subvector of $\vec{x}$ starting at element $i$ and including all elements up to (including) element $j$.
Furthermore $\min \vec{z}$ and $\max \vec{z}$ mean the minimum or maximum element of a vector $\vec{z}$. 
The transformed system should be tight in the sense that given a subvector $\vec{x}_{M-s \,\dots\, M}$ which satisfies the first $s+1$ inequalities there must exist remaining elements $\vec{x}_{1 \,\dots\, M-s-1}$ so that $\vec{x}$ satisfies all inequalities.
This is equivalent to demanding that the transformed inequalities must not allow values for an element $x_i$ so that the ranges of allowed values for other elements of $\vec{x}$ becomes empty.
Obviously the matrices $L^i$, $\widehat{L}^i$, $H^i$ and $\widehat{H}^i$ depend on $A$ and must be determined.

Multiplying an inequality by a positive, non-zero factor will result in an equivalent system, where equivalent means that it has exactly the same set of solutions as the original system.
Thus, by dividing each line $i$ with $A_{i1} \neq 0$ by the factor $\abs{A_{i1}}$ and rearranging, we can bring a system of the form \eqref{eq:ieq_orignal} into the equivalent form
\begin{align}
x_1  + \sum_{j=2}^M D_{hj} x_j  & \geq d_h \,, \quad h \in \{1, \dots, H \} \label{eq:ieq_tf_plus}  \\
-x_1 + \sum_{j=2}^M E_{kj} x_j  & \geq e_k \,, \quad k \in \{1, \dots, K \} \label{eq:ieq_tf_minus} \\
       \sum_{j=2}^M F_{lj} x_j  & \geq f_l \,, \quad l \in \{1, \dots, L \} \label{eq:ieq_tf_zero} 
\end{align}
with $H + K + L = N$.
It is clear that adding two inequalities will not reduce the set of solutions, \ie if $\vec{x}$ is a solution to the inequalities $\vec{a}^T \vec{x} \geq \alpha$ and $\vec{b}^T \vec{x} \geq \beta$, then $\vec{x}$ is also a solution to the inequality $(\vec{a} + \vec{b})^T \vec{x} \geq \alpha + \beta$.
Consequently by adding each inequality from \eqref{eq:ieq_tf_plus} to each inequality from \eqref{eq:ieq_tf_minus} and dropping the used inequalities we arrive at the reduced system with $x_1$ eliminated,
\begin{alignat}{2}
\sum_{j=2}^M (D_{hj} + E_{kj}) \, x_j  & \geq d_h + e_k \,, \quad && h \in \{1, \dots, H \},\, k \in \{1, \dots, K \} \,, \label{eq:ieq_tf_eli1}  \\
\sum_{j=2}^M F_{lj} x_j                & \geq f_l \,, \quad       && l \in \{1, \dots, L \} \label{eq:ieq_tf_eli2}  \,,
\end{alignat}
which has at least the solutions $\vec{x}$ of the original system consisting of \cref{eq:ieq_tf_plus,eq:ieq_tf_minus,eq:ieq_tf_zero}.
Fourier and Motzkin \cite{dantzig1973fourier} observed that both system are indeed equivalent.
To verify this, we have to show that for each solution $\vec{x}_{2 \cdots M}$ of \cref{eq:ieq_tf_eli1,eq:ieq_tf_eli2}, there exists $x_1$ so that the combined $\vec{x}$ satisfies \cref{eq:ieq_tf_plus,eq:ieq_tf_minus,eq:ieq_tf_zero}.
From \eqref{eq:ieq_tf_plus} and \eqref{eq:ieq_tf_minus} we see that an $x_1$ satisfying the original system is given by
\begin{equation}
\min_k \left(\sum_{j=2}^M E_{kj} x_j - e_k \right) \geq x_1 \geq \max_h \left(-\sum_{j=2}^M D_{hj} x_j + d_h \right)
\end{equation}
and rewriting \eqref{eq:ieq_tf_eli1} as
\begin{equation}
\sum_{j=2}^M E_{kj} x_j - e_k \geq -\sum_{j=2}^M D_{hj} x_j + d_h \,, \quad  h \in \{1, \dots, H \},\, k \in \{1, \dots, K \}
\end{equation}
shows that an $x_1$ with this property exists if the reduced system is satisfied.

By iteratively applying the reduction method just described, we can sequentially eliminate $x_1$, $x_2$ and so on up to $x_M$, as long as there exists at least one pair of inequalities with opposite signs for a specific $x_i$.
If this is not the case, then the remaining $x_{i+1 \,\dots\, M}$ are not affected by these inequalities since a value for $x_i$ can always be found after determining $x_{i+1 \,\dots\, M}$ because $x_i$ is bounded from one side only; consequently when $x_i$ occurs with positive or negative sign only, all inequalities containing $x_i$ can be dropped to progress with the elimination.
After $x_M$ has been eliminated, what remains is a system of constant inequalities of the form
\begin{equation}
0 \geq f_l \,, \quad  l \in \{1, \dots, L \} \label{eq:ieq_tf_feasible}  \,.
\end{equation}
If these inequalities contain a contradiction, \ie if any $f_l$ is positive, the original system of inequalities is inconsistent and the set of solutions for $\vec{x}$ is empty.

This elimination method gives rise to \cref{alg:fouriermotzkin} which has been adapted from \cite{dantzig2016linear,dantzig2006linear1,dantzig2006linear2} to work on matrix $A$ only and thus solving the system of inequalities for arbitrary $\vec{b}$.
The algorithm produces matrices $L^i$, $H^i$ and $\widehat{L}^i$, $\widehat{H}^i$ for $i \in \{1, \dots, M\}$ that can be inserted into the inequalities \eqref{eq:ieq_tf_m} to \eqref{eq:ieq_tf_1} to subsequently obtain the ranges for each element of $\vec{x}$.
It also outputs the feasibility matrix $F$, with the property that if $F \vec{b} \leq \vec{0}$, then there exist a solution for a particular $\vec{b}$.

\begin{algorithm}[tbp]\DontPrintSemicolon
\caption{Fourier-Motzkin elimination for a system of linear inequalities $A \, \vec{x} \geq \vec{b}$}\label{alg:fouriermotzkin}
\KwIn{matrix $A \in \setR^{N \times M}$}
\KwOut{matrices $L^i$, $H^i$ and $\widehat{L}^i$, $\widehat{H}^i$ for $i \in \{1, \dots, M\}$ for use in \eqref{eq:ieq_tf_m} to \eqref{eq:ieq_tf_1}; feasibility matrix $F$}
\vspace{2mm}
$B \longleftarrow \idmatrix_N$		\tcp*{initialize $B$ with identity matrix} 
\For(\tcp*[f]{loop over variables to eliminate}){$k \in \{1, \dots, M\}$}{	
	\tcp{divide each row i by $\abs{A_{ik}}$}	
	\For{$i \in \{1, \dots N\}$}{
		\If{$A_{ik} \neq 0$}{
			$A_{i\cdot} \longleftarrow \frac{1}{\abs{A_{ik}}} \, A_{i\cdot}$ \\
			$B_{i\cdot} \longleftarrow \frac{1}{\abs{A_{ik}}} \, B_{i\cdot}$
		}
	}
	\vspace{2mm}
	\tcp{extract solution matrices}
	$\zeta \longleftarrow \{i \in \setZ \setbar A_{ik} = 0  \}$;
	$\phi  \longleftarrow \{i \in \setZ \setbar A_{ik} = +1 \}$;
	$\mu   \longleftarrow \{i \in \setZ \setbar A_{ik} = -1 \}$ \\			
	$S \longleftarrow -$ columns $\{k+1, \dots, M\}$ of $A$ \\
	$L^k \longleftarrow $ rows $\phi$ of $B$;
	$H^k \longleftarrow -$ rows $\mu$ of $B$ \\
	$\widehat{L}^k \longleftarrow $ rows $\phi$ of $S$;		
	$\widehat{H}^k \longleftarrow -$ rows $\mu$ of $S$ \\			
	\vspace{2mm}
	\tcp{eliminate $x_k$}
	\If{$\phi = \emptyset \,\land\, \mu = \emptyset$}{
		\tcp{$x_k$ does not occur, nothing to eliminate}
	}\ElseIf{$\phi = \emptyset \,\lor\, \mu = \emptyset$}{
		\tcp{$x_k$ occurs with coefficient $+1$ or $-1$ only}
		$A \longleftarrow $ rows $\zeta$ of $A$;
		$B \longleftarrow $ rows $\zeta$ of $B$  
	}\Else{
		\tcp{$x_k$ occurs with coefficients $+1$ and $-1$}
		$A' \longleftarrow $ rows $\zeta$ of $A$;
		$B' \longleftarrow $ rows $\zeta$ of $B$  \\
		\For{$p \in \phi$}{
			\For{$n \in \mu$}{
				$A' \longleftarrow A'$ with row $A_{p\cdot} + A_{n\cdot}$ appended \\
				$B' \longleftarrow B'$ with row $B_{p\cdot} + B_{n\cdot}$ appended 					
			}			
		}
		$A \longleftarrow A'$; $B \longleftarrow B'$
	}
}
$F \longleftarrow B$ \tcp*{inequalities with no variables left}
\end{algorithm}

\clearpage
\section{Elementwise-defined Functions and their Derivatives}\label{sec:ediff}
We introduce the situations that can occur when calculating the derivatives of elementwise-defined tensors using the following set of examples.
Then we will describe a general method to derive expressions for the derivatives of elementwise-defined tensors, where the indices of the arguments are an arbitrary linear combination of the indices of the function output.
Summations within these expressions are allowed.

Consider the vector-valued function $\vec{f}^1: \setR^N \to \setR^N$, that is defined by specifying how each element of $\vec{f}^1(\vec{x})$ depends on the elements of its arguments $\vec{x}$.
For example, a very simple example for such a function is
\[ f^1_i(\vec{x}) = \sin x_i \,. \]
Here it is straightforward to see that its Jacobian is given by
\[ \frac{\partial f^1_i}{\partial x_{i'}} = \delta_{i,i'} \, \cos x_{i'} \]
since element $i$ of $\vec{f}^1$ only depends by element $i$ of its arguments $\vec{x}$.
Hence, the Kronecker delta was introduced in the above expression to make sure that $\partial f^1_i / \partial x_{i'} =0$ for $i \neq i'$.

Further assume that $\vec{f}$ is part of a scalar function $l$ with $l(\vec{x}) = g(\vec{f}(\vec{x}))$ and the derivatives of $l$ \wrt the elements of $\vec{x}$ are to be derived.
The derivatives $\partial g / \partial f_i$ are supposed to be known.
Let us introduce the notation
\[ \d \bullet_{\alpha} = \frac{\partial l}{\partial \bullet_{\alpha}} \]
for the derivatives of $l$ \wrt an element of a variable or function.
In the context of deep learning this is the derivative we are usually interested in, since it provides the gradient of a loss function $l$ and is thus used for minimization of the loss.
The explicit computation of the Jacobians $\partial f_i / \partial x_j$ is usually not of interest since it wastes space.

We obtain for our function $\vec{f}^1(\vec{x})$,
\[ \d f^1_{i'} = \sum_i \frac{\partial g}{\partial f^1_{i}}\, \frac{\partial f^1_i}{\partial x_{i'}} = \sum_i \d g_i \, \delta_{i,i'} \, \cos x_i = \d g_{i'} \, \cos x_{i'}  \,. \]
Let us now consider a slightly more complicated example given by the function $f^2: \setR^N \times \setR^{N \times N} \to \setR^{N \times N}$ of two arguments with the element-wise specification
\[ f^2_{ij}(\vec{x}, y) = x_i \, y_{ij} \,. \]
The (extended) Jacobians \wrt $x$ and $y$ are given by
\[ \frac{\partial f^2_{ij}}{\partial x_{i'}}   = \delta_{i,i'} \, y_{i'j} \,, \quad\quad
\frac{\partial f^2_{ij}}{\partial y_{i'j'}} = \delta_{i,i'} \, \delta_{j,j'} \, x_{i'} \,, \]
where the derivative \wrt $x$ does not contain a Kronecker delta for index $j$, since it is not used to index variable $x$.
Consequently application of the chain rule gives the following derivatives of $l$,
\[ \d x_{i'}    = \sum_j \d g_{i'j} \, y_{i'j}  \,, \quad\quad
\d y_{i'j'}  = \d g_{i'j'} \, x_{i'} \,, \]
where the lack of index $j$ on variable $x$ has lead to a summation over this index.
Another situation is demonstrated by the function $f^3: \setR^{N \times N} \to \setR^N$ with
\[ f^3_{i}(x) = x_{ii}^3 \,. \]
The Jacobian, 
\[ \frac{\partial f^3_i}{\partial x_{i'j'}} = \delta_{i,i'} \, \delta_{i,j'} \, 3 x_{i'j'}^2  \,, \]
now contains two Kronecker deltas for the index $i$ to express that $i=i'=j'$ must hold so that the derivative is non-zero.
This leads to the derivative of $l$,
\[ \d x_{i'j'} = \delta_{i',j'} \, \d g_{i'} \, 3 x_{i'j'}^2 \,, \]
which now contains a Kronecker delta itself, since it has not been canceled out by a corresponding summation.
A good example for a function containing a summation over its arguments is the matrix dot product,
\[ f^4_{ij}(x, y) = \sum_k x_{ik} \, y_{kj} \,, \]
which has the (extended) Jacobians
\[ \frac{\partial f^4_{ij}}{\partial x_{i'k'}} = \delta_{i,i'} \, \sum_k \delta_{k,k'} \, y_{kj}  \,, \quad\quad
\frac{\partial f^4_{ij}}{\partial y_{k'j'}} = \delta_{j,j'} \, \sum_k \delta_{k,k'} \, x_{ik}  \,. \]
Thus the derivatives of $l$ evaluate to
\begin{align*}
\d x_{i'k'} &= \sum_i \sum_j \d g_{ij} \, \delta_{i,i'} \, \sum_k \delta_{k,k'} \, y_{kj} = \sum_j \d g_{i'j} \, y_{k'j} \,, \\
\d y_{k'j'} &= \sum_i \sum_j \d g_{ij} \, \delta_{j,j'} \, \sum_k \delta_{k,k'} \, x_{ik} = \sum_i \d g_{ij'} \, x_{i k'} \,.
\end{align*}
Note that the summation indices of the derivatives have changed over from $k$ to $j$ and $i$ respectively.
Finally consider the situation where the indices of the argument are given by a linear combination of the function indices, as demonstrated by $f^5: \setR^{N \times M} \to \setR^{N M}$ with
\[ f^5_{ij}(\vec{x}) = \exp x_{Mi + j} \,. \]
Its Jacobian is straightforward to express,
\[ \frac{\partial f^5_{ij}}{\partial x_{i'}} = \delta_{Mi + j,i'} \, \exp x_{Mi + j}  \,, \]
however to efficiently express the derivative of $l$,
\[ \d x_{i'} = \sum_i \sum_j \d g_{ij} \, \delta_{Mi + j,i'} \, \exp x_{Mi + j}  \,, \]
the occurring Kronecker delta should be combined with one of the sums, because one of them is redundant.
To do so it is necessary to solve the equation $Mi +j =i'$ for $j$, which is trivial in this example.
The solution is given by $j = i' - Mi$ and after substitution this results in
\[ \d x_{i'} = \sum_i \d g_{i,i'-Mi} \, \exp x_{i'}  \,. \]
Note that the sum range must be chosen appropriately, which is not shown here.
We have seen that, depending on the constellation of indices of the arguments of a elementwise-defined function, the derivative will either introduce additional summations, drop existing summations, introduce Kronecker deltas or even require substitution of the solution of a linear equation system into the indices or a combination of these things.

\subsection{Computing element-wise derivative expressions}
We first describe the method without accounting for summations inside the function and reintroduce them later.
Generally the problem of computing expressions for elementwise derivatives can be stated as follows.
Let $\vec{\alpha} = (\alpha_1, \alpha_2, \dots, \alpha_{D_f})$ be a multi-index and let the tensor-valued function $f: \setR^{N^1_{1} \times \cdots \times N^1_{D_1}} \times \cdots \times \setR^{N^P_{1} \times \cdots \times N^P_{D_P}} \to \setR^{N^f_1 \times \cdots \times N^f_{D_f}}$ taking $P$ tensor arguments called $x^1, x^2, \dots, x^P$ be specified element-wise, 
\begin{equation}
f_{\vec{\alpha}} (x^1, x^2, \dots, x^P) = \fbar (x^1_{A^1 \vec{\alpha}}, x^2_{A^2 \vec{\alpha}}, \dots, x^P_{A^P \vec{\alpha}} ) \,,
\end{equation}
where each matrix $A^p: \setZ^{D_f} \to \setZ^{D_p}$ maps from the indices of $f$ to the indices of its argument $x^p$.
Such a linear transform covers all the cases shown in the introductory examples.
If the same argument $x^p$ should appear multiple times with different indices, we shall treat it as different arguments (by renaming the different occurrences) and sum over the resulting expressions for the derivatives after they have been obtained.
Note that $\fbar: \setR \times \cdots \times \setR \to \setR$ is a \emph{scalar} function.
Furthermore let $g: \setR^{N^f_1 \times \cdots \times N^f_{D_f}} \to \setR$ be a scalar-valued function and let $l = g \circ f$.
Let $\d f \in \setR^{N^f_1 \times \cdots \times N^f_{D_f}}$ be the tensor of derivatives of $l$ \wrt the elements of $f$, thus by above definition
\begin{equation}
\d f_{\vec{\alpha}} = \frac{\partial l}{\partial f_{\vec{\alpha}}} = \frac{\partial g}{\partial f_{\vec{\alpha}}} \,.
\end{equation}
The objective is to obtain expressions that specify the derivatives of $l$ \wrt the elements of each $x^p$ element-wise, \ie
\begin{equation}
\d x^p_{\vec{\beta}^p} = \frac{\partial l}{\partial x^p_{\vec{\beta}^p}} \label{eq:ewd_dxp_beta}
\end{equation}
where $\vec{\beta}^p = (\beta^p_1, \beta^p_2, \dots, \beta^p_{D_p})$ is a multi-index enumerating the elements of $x^p$. 

Applying the chain rule to \eqref{eq:ewd_dxp_beta} gives
\begin{equation}\label{eq:ewd_dxp_beta_2}
\d x^p_{\vec{\beta}^p} = \frac{\partial l}{\partial x^p_{\vec{\beta}^p}} = \sum_{\substack{\vec{1} \leq \vec{\alpha} \leq \vec{N}^f \\ A^p \vec{\alpha} = \vec{\beta}^p }} \frac{\partial l}{\partial f_{\vec{\alpha}}} \frac{\partial f_{\vec{\alpha}}}{\partial x^p_{\vec{\beta}^p}} = \sum_{\vec{\alpha} \in \Gamma(\vec{\beta}^p)} \d f_{\vec{\alpha}} \, \frac{\partial \fbar}{\partial x^p_{A^p \vec{\alpha}}} \,
\end{equation}
and since $\fbar$ is a scalar function, computing the scalar derivative $\partial \fbar / \partial x^p_{A^p \vec{\alpha}}$ is straightforward using the strategy described in \cref{sec:impl_autodiff_graph}.
Thus the main challenge is to efficiently evaluate the summation over the set
\begin{equation}\label{eq:ewd_gamma}
\Gamma(\vec{\beta}^p) = \{\vec{\alpha} \in \setZ^{D_f} \setbar \vec{1} \leq \vec{\alpha} \leq \vec{N}^f \,\land\, A^p \vec{\alpha} = \vec{\beta}^p  \} \,,
\end{equation}
\ie find all \emph{integer} vectors $\vec{\alpha}$ that fulfill the relation $A^p \vec{\alpha} = \vec{\beta}^p$ and lie within the range $\vec{1} \leq \vec{\alpha} \leq \vec{N}^f$ determined by the shape of $f$.

An elementary approach, as demonstrated in the introductory examples, is to rewrite \cref{eq:ewd_dxp_beta_2} as
\begin{equation}\label{eq:ewd_dxp_beta_3}
\d x^p_{\vec{\beta}^p} = \sum_{\alpha_1=1}^{N^1} \cdots \sum_{\alpha_{D_f}=1}^{N^{D_f}} \delta_{A^p \vec{\alpha} - \vec{\beta}^p} \, \d f_{\vec{\alpha}} \, \frac{\partial \fbar}{\partial x^p_{A^p \vec{\alpha}}}
\end{equation}
where the single-argument Kronecker delta is given by $\delta_{\vec{t}} = 0$ for $\vec{t} \neq \vec{0}$ and $\delta_{\vec{0}} = 1$.
Thus for each index $\vec{\alpha}$ of $f$ we test explicitly if it contributes to the derivative of index $\vec{\beta}^p$ of argument $x^p$ and if so, we include that element in the summation.
By evaluating \eqref{eq:ewd_dxp_beta_3} for all $\vec{\beta}^p$ in parallel the cost of iterating over $\vec{\alpha}$ can be amortized over all elements of $\d x^p$.
However, if multiple threads are used to perform this iteration, as it is required to gain acceptable performance on modern GPUs, locking becomes necessary to serialize writes to the same element of $\d x^p$.
If $A^p$ has low rank, write collisions on $\d x^p_{A^p \vec{\alpha}}$ become likely, leading to serialization and thus considerable performance loss.\footnote{The CUDA programming guide \cite{cuda} is vague about the performance penalties associated with atomic addition to the same memory access from within multiple threads. Nonetheless, experiments \cite{cuda_atomic1,cuda_atomic2} show that performance can be degraded by up to a factor of 32 due to locking and resulting serialization.}
Another drawback of this approach is that even if only a subset of elements of the derivative $\d x^p$ are required, the summation must always be performed over the whole range of $\vec{\alpha}$.
Furthermore, while not being of great importance for minimization of loss functions in machine learning, it is regrettable that no symbolic expression for $\d x^p_{\vec{\beta}^p}$ is obtained using this method.

For these reasons it is advantageous to find a form of the set \cref{eq:ewd_gamma} that directly enumerates all $\vec{\alpha}$ belonging to a particular $\vec{\beta}^p$.
This requires solving $A^p \vec{\alpha} = \vec{\beta}^p$ for $\vec{\alpha}$.
In general, a set of linear equations with integer coefficients over integer variables, has either none, one or infinitely many solutions.
The set of solutions can be fully described using the pseudo-inverse, cokernel and kernel.
Thus, let $I$ be the pseudo-inverse, $C$ the cokernel and $K$ the kernel of the integer matrix $A$ as defined in \cref{sec:integer_eqs}.
Using these matrices we can rewrite \eqref{eq:ewd_gamma} as
\begin{equation}\label{eq:ewd_gamma_exp1}
\Gamma(\vec{\beta}^p) = \{I \vec{\beta}^p + K \vec{z} \setbar C \vec{\beta}^p = \vec{0} \,\land\, I \vec{\beta}^p \in \setZ^{D_f} \,\land\, \vec{z} \in \setZ^\kappa \,\land\, \vec{1} \leq I \vec{\beta}^p + K \vec{z} \leq \vec{N}^f \} \,,
\end{equation}
where $\kappa$ is the dimensionality of the kernel of $A$.
The conditions $C \vec{\beta}^p = \vec{0}$ and $I \vec{\beta}^p \in \setZ^{D_f}$ determine whether the set is empty or not for a particular $\vec{\beta}^p$ and since they are independent of $\vec{z}$, they only need to be checked once for each $\vec{\beta}^p$.
Thus if these conditions do not hold, we can immediately conclude that $\d x^p_{\vec{\beta}^p} = 0$.
Otherwise, in order to further simplify the set specification, we need to find the elements of the set
\begin{equation}\label{eq:ewd_sigma}
\Sigma(\vec{\beta}^p) = \{ \vec{z} \in \setZ^\kappa \setbar \vec{1} \leq I \vec{\beta}^p + K \vec{z} \leq \vec{N}^f \}
\end{equation}
containing all $\vec{z}$ that generate values for $\vec{\alpha}$ within its valid range.
Since $\vec{\alpha}(\vec{z}) = I \vec{\beta}^p + K \vec{z}$ is an affine transformation, the set $\Sigma(\vec{\beta}^p)$ must be convex.
By rewriting the system of inequalities defining the set $\Sigma(\vec{\beta}^p)$ as
\begin{align}
K\,\vec{z}  & \geq \vec{1}    - I \vec{\beta}^p \\
-K\,\vec{z} & \geq -\vec{N}^f + I \vec{\beta}^p
\end{align}
we can apply the Fourier-Motzkin algorithm described in \cref{sec:lin_ieq} to obtain the boundaries of the convex set in closed form.
The Fourier-Motzkin algorithm produces matrix $L^i$, $H^i$ and $\widehat{L}^i$, $\widehat{H}^i$ so that \eqref{eq:ewd_sigma} can be written as
\begin{align}
\Sigma(\vec{\beta}^p) = \{ \vec{z} \in \setZ^\kappa \setbar 
& \ceil{\max (L^\kappa \vec{b})} \leq z_\kappa \leq \floor{\min (H^\kappa \vec{b})} \,\land\, \nonumber \\
& \ceil{\max (L^{\kappa-1} \vec{b} + \widehat{L}^{\kappa-1} \vec{z}_{\kappa})} \leq z_{\kappa-1} \leq \floor{\min (H^{\kappa-1} \vec{b} + \widehat{H}^{\kappa-1} \vec{z}_{\kappa})} \,\land\, \nonumber \\
& \cdots \,\land\, \nonumber \\
& \ceil{\max (L^{1} \vec{b} + \widehat{L}^{1} \vec{z}_{2 \,\dots\, \kappa})} \leq z_{1} \leq \floor{\min (H^{1} \vec{b} + \widehat{H}^{1} \vec{z}_{2 \,\dots\, \kappa})} \}  \label{eq:ewd_sigma2}
\end{align}
where
\[ \vec{b}(\vec{\beta}^p) \teq \begin{bmatrix} \vec{1}    - I \vec{\beta}^p  \\ -\vec{N}^f + I \vec{\beta}^p \end{bmatrix}  \, \]
and $\floor{\bullet}$ and $\ceil{\bullet}$ are the floor and ceiling respectively.
Since the Fourier-Motzkin algorithm executes independently of the value of $\vec{b}$, the computationally intensive procedure of computing the matrices $L^i$, $H^i$ and $\widehat{L}^i$, $\widehat{H}^i$ is only done once for each kernel matrix $K$.
Afterwards, computing the boundaries for a particular index $\vec{\beta}^p$ requires only four matrix multiplications per dimension and the determination of the minimum and maximum value of a vector.

\begin{figure}[tbp]
\centering
\begin{subfigure}{0.48\textwidth}
\centering
\resizebox{!}{5.8cm}{\begin{tikzpicture}
    \tikzstyle{op}=[circle,blue,very thick,draw,minimum size=17pt,inner sep=0pt,text=black]
    \tikzstyle{deriv}=[rectangle,red,very thick,draw,minimum size=17pt,inner sep=0pt,text=black,fill]
    \tikzstyle{diff}=[circle,red,very thick,draw,minimum size=17pt,inner sep=0pt,text=black]
	\tikzstyle{annot} = [text width=4em, text centered]	
	\tikzstyle{p} = [thick,->,blue]
	\tikzstyle{pd} = [thick,->,red]
	\tikzstyle{el} = [very thin,shorten >=-#1,shorten <=-#1]

	\def\r{-1.0} \def\c{-0.7} \def\dr{-4.0} \def\drr{4.8}

	\clip(-0.8,-0.5) rectangle (7.8,5.8);

	\draw[very thin,->] (0,0) -- (7.5,0);
	\draw[very thin,->] (0,0) -- (0,5.5);
	\draw[very thin] (0,0) circle (0.15);
	\draw[very thin,fill] (0,0) circle (0.05);

	\foreach \x in {0,1,2,4,5,6,7}
    	\draw (\x cm,1pt) -- (\x cm,-3pt) node[anchor=north] {$\x$};
	\foreach \y in {1,2,3,4,5}
    	\draw (1pt,\y cm) -- (-1pt,\y cm) node[anchor=east] {$\y$};

	\foreach \x in {1,3,5,7}
		\foreach \y in {0,1,2,3,4,5}
    		\node[gray] at (\x,\y) {\textbullet};

    \node[] at (7.6,-0.3) {$\alpha_1$};
    \node[] at (-0.3,5.4) {$\alpha_2$};
    \node[] at (-0.4,0.0) {$\alpha_3$};

	\draw[very thick] (0,0) rectangle (6, 4);

	\draw[line width=0.5mm,->] (3,0) -- (5,1);
	\node at (3.8,0.75) {$K_{\bullet 1}$};

	\draw[line width=0.5mm,->] (3,0) -- (3,1);
	\node at (2.6,0.75) {$K_{\bullet 2}$};

	\node at (1,1) {\textbullet};
	\node at (1,2) {\textbullet};
	\node at (1,3) {\textbullet};
	\node at (3,1) {\textbullet};
	\node at (3,2) {\textbullet};
	\node at (3,3) {\textbullet};
	\node at (5,1) {\textbullet};
	\node at (5,2) {\textbullet};
	\node at (5,3) {\textbullet};	
	
	\node at (3,-0.35) {$I \boldsymbol{\beta}$};	
	
	
	\node[red] at (1,0) {\textbullet};
	\node[red] at (3,0) {\textbullet};	
	\node[red] at (5,0) {\textbullet};	
	\node[red] at (1,4) {\textbullet};		
	\node[red] at (3,4) {\textbullet};			
	\node[red] at (5,4) {\textbullet};		
	
	\draw[red,very thick] (0,0) rectangle (6, 4);

\end{tikzpicture}}
\caption{in $\vec{\alpha}$-space}
\end{subfigure}\hfill
\raisebox{0.4cm}{\begin{subfigure}{0.48\textwidth}
\centering
\resizebox{!}{6.7cm}{\begin{tikzpicture}
    \tikzstyle{op}=[circle,blue,very thick,draw,minimum size=17pt,inner sep=0pt,text=black]
    \tikzstyle{deriv}=[rectangle,red,very thick,draw,minimum size=17pt,inner sep=0pt,text=black,fill]
    \tikzstyle{diff}=[circle,red,very thick,draw,minimum size=17pt,inner sep=0pt,text=black]
	\tikzstyle{annot} = [text width=4em, text centered]	
	\tikzstyle{p} = [thick,->,blue]
	\tikzstyle{pd} = [thick,->,red]
	\tikzstyle{el} = [very thin,shorten >=-#1,shorten <=-#1]

	\def\r{-1.0} \def\c{-0.7} \def\dr{-4.0} \def\drr{4.8}

	\clip(-2.5,-1.5) rectangle (2.8,5.8);

	\draw[very thin,->] (-2.5,0) -- (2.8,0);
	\draw[very thin,->] (0,-1.5) -- (0,5.8);

    \node[] at (2.7,-0.3) {$z_1$};
    \node[] at (-0.3,5.6) {$z_2$};

	\foreach \x in {-2,-1,0,1,2}
    	\draw (\x cm,1pt) -- (\x cm,-1pt) node[anchor=north] {$\x$};
	\foreach \y in {-1,1,2,3,4,5}
    	\draw (1pt,\y cm) -- (-1pt,\y cm) node[anchor=east] {$\y$};

	\foreach \x in {-2,-1,0,1,2}
		\foreach \y in {-1,0,1,2,3,4,5}
    		\node[gray] at (\x,\y) {\textbullet};
	
	\node[red] at (-1,1) {\textbullet};	
	\node[black] at (-1,2) {\textbullet};	
	\node[black] at (-1,3) {\textbullet};	
	\node[black] at (-1,4) {\textbullet};	
	\node[red] at (-1,5) {\textbullet};	
	\node[red] at (0,0) {\textbullet};	
	\node[black] at (0,1) {\textbullet};	
	\node[black] at (0,2) {\textbullet};	
	\node[black] at (0,3) {\textbullet};	
	\node[red] at (0,4) {\textbullet};	
	\node[red] at (1,-1) {\textbullet};	
	\node[black] at (1,0) {\textbullet};	
	\node[black] at (1,1) {\textbullet};	
	\node[black] at (1,2) {\textbullet};	
	\node[red] at (1,3) {\textbullet};

	\draw[red,very thick] (-1.5,5.5) -- (1.5,2.5) -- (1.5,-1.5) -- (-1.5,1.5) -- (-1.5,5.5);

\end{tikzpicture}}
\caption{in $\vec{z}$-space}
\end{subfigure}}
\caption[A one-dimensional parameter index $\vec{\beta}$ driven by a three-dimensional function index $\vec{\alpha}$.]{
A one-dimensional parameter index $\vec{\beta}$ driven by a three-dimensional function index $\vec{\alpha}$.
(a) This shows $\vec{\alpha}$-space as a cut through the $\alpha_1$-$\alpha_2$ plane, \ie the $\alpha_3$-axis is perpendicular to this drawing.
The one-dimensional index of $x$ is given by $\vec{\beta} = A \vec{\alpha}$ with $A = \begin{pmatrix} 1 & -2 & -2\end{pmatrix}$.
This yields $\vec{\alpha} = I \vec{\beta} + K \vec{z}$ with the pseudo-inverse $I = \begin{pmatrix} 1 \\ 0 \\ 0 \end{pmatrix}$.
A possible choice for the two-dimensional kernel is $K = \begin{pmatrix} 2 & 0 \\ 1 & 1 \\ 2 & 1 \end{pmatrix}$.
For $\vec{\beta} = \begin{pmatrix} 3 \end{pmatrix}$ the set of possible values for $\vec{\alpha}$ is given by the sum of $I\vec{\beta}$ and integer linear combinations of the columns of the kernel matrix $K$.
The constraint \eqref{eq:ewd_sigma} imposed by the range of $\vec{\alpha}$ requires valid values to lie inside the red rectangle.
(b) By mapping this rectangle into the domain of the kernel, \ie $\vec{z}$-space, we obtain a parallelogram.
Thus values for $\vec{z}$ as allowed by \eqref{eq:ewd_sigma2} are the integer points that lie within this parallelogram, \ie 
$\Sigma(\begin{pmatrix} 3 \end{pmatrix}) = \{ z \in \setZ \setbar -1 \leq z_2 \leq 5 \,\land\, \max(-1, -z_2) \leq z_1 \leq \min(1, 4-z_2) \}$ corresponding to the 15 points inside the rectangle in $\vec{\alpha}$-space.
This causes the range of $z_1$ to become dependent on the value of $z_2$.
}
\label{fig:deriv_lattice2}
\end{figure}
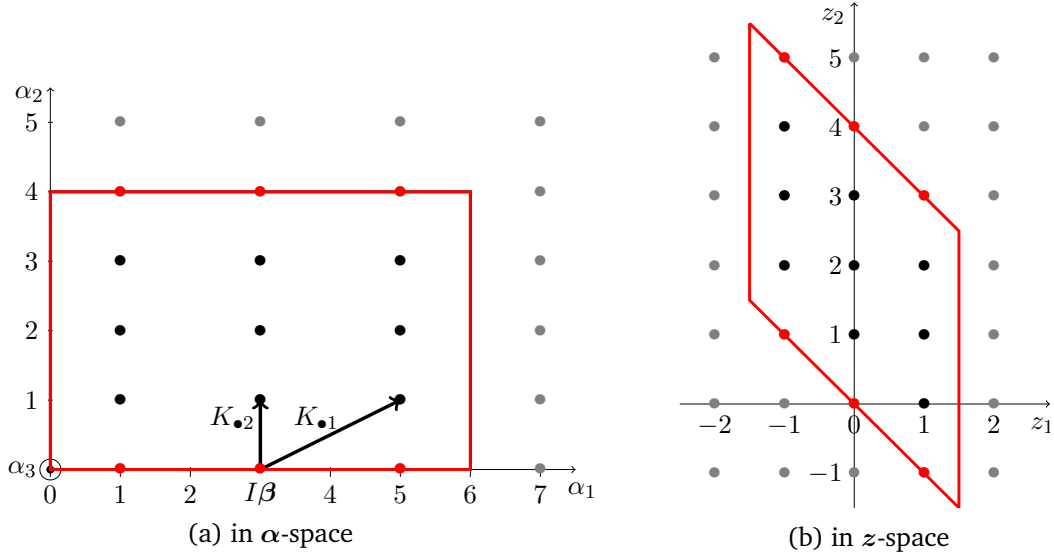

An example for a one-dimensional kernel, \ie line, is shown in \cref{fig:deriv_lattice1}. 
In this case \eqref{eq:ewd_sigma2} consists of only one condition for $z_1$ and describes the part of the line that is inside the range specified by $\vec{1} \leq \vec{\alpha} \leq \vec{N}^f$.
Another example, this time for a two-dimensional kernel, \ie plane, is shown in \cref{fig:deriv_lattice2}.
Due to the choice of the kernel basis, the range specified by $\vec{1} \leq \vec{\alpha} \leq \vec{N}^f$ becomes a parallelogram in the domain of the kernel and thus the resulting ranges for $z_1$ and $z_2$ are dependent on each other.
 
\begin{figure}[tb]
\centering
\begin{tikzpicture}
    \tikzstyle{op}=[circle,blue,very thick,draw,minimum size=17pt,inner sep=0pt,text=black]
    \tikzstyle{deriv}=[rectangle,red,very thick,draw,minimum size=17pt,inner sep=0pt,text=black,fill]
    \tikzstyle{diff}=[circle,red,very thick,draw,minimum size=17pt,inner sep=0pt,text=black]
	\tikzstyle{annot} = [text width=4em, text centered]	
	\tikzstyle{p} = [thick,->,blue]
	\tikzstyle{pd} = [thick,->,red]
	\tikzstyle{el} = [very thin,shorten >=-#1,shorten <=-#1]

	\def\r{-1.0} \def\c{-0.7} \def\dr{-4.0} \def\drr{4.8}

	\clip(-1.5,-0.5) rectangle (7.8,5.7);

	\draw[very thin,->] (-1.5,0) -- (7.5,0);
	\draw[very thin,->] (0,0) -- (0,5.5);
	
    \node[] at (7.6,-0.3) {$\alpha_1$};
    \node[] at (-0.3,5.4) {$\alpha_2$};	

	\foreach \x in {-1,0,1,2,3,4,5,6,7}
    	\draw (\x cm,1pt) -- (\x cm,-1pt) node[anchor=north] {$\x$};
	\foreach \y in {1,2,3,4,5}
    	\draw (1pt,\y cm) -- (-1pt,\y cm) node[anchor=east] {$\y$};

    \node[] at (1,1) {\textbullet};
    \node[] at (3,2) {\textbullet};
    \node[] at (5,3) {\textbullet};
    \node[gray] at (7,4) {\textbullet};

	\draw[very thick] (0,0) rectangle (6, 4);

	\draw[el=20cm] (1,1) -- (3,2);

	\draw[thick,->] (-1,0) -- (1,1);
	\node at (0.6,1.05) {$\mathbf{K}$};

	\node[gray] at (-1,0) {\textbullet};
	\node at (-1.1,0.3) {$I \boldsymbol{\beta}$};	
	
	\node[red] at (0,0.5) {$\times$};
	\node at (0.4,0.5) {$\boldsymbol{l}$};
	
	\node[red] at (6,3.5) {$\times$};
	\node at (5.6,3.5) {$\boldsymbol{h}$};

\end{tikzpicture}
\caption[A one-dimensional parameter index $\vec{\beta}$ driven by a two-dimensional function index $\vec{\alpha}$.]{
A one-dimensional parameter index $\vec{\beta}$ driven by a two-dimensional function index $\vec{\alpha}$ shown in $\vec{\alpha}$-space.
The one-dimensional index of $x$ is given by $\vec{\beta} = A \vec{\alpha}$ with $A = \begin{pmatrix} 1 & -2 \end{pmatrix}$.
This yields $\vec{\alpha} = I \vec{\beta} + K \vec{z}$ with the pseudo-inverse $I^T = \begin{pmatrix} 1 & 0 \end{pmatrix}$ and one-dimensional kernel $K^T = \begin{pmatrix} 2 & 1 \end{pmatrix}$.
For $\vec{\beta} = \begin{pmatrix} -1 \end{pmatrix}$ the set of possible values for $\vec{\alpha}$ lies on the marked line with direction vector given by the kernel $K$.
This set is limited by the requirement that $\vec{\alpha}$ must be integer, thus only the marked points on the line are valid values for $\vec{\alpha}$.
Furthermore the constraint \eqref{eq:ewd_sigma} imposed by the range of $\vec{\alpha}$ requires valid values to lie between the points marked $\vec{l}$ and $\vec{h}$.
Thus values for $z$ as allowed by \eqref{eq:ewd_sigma2} are $\Sigma(\begin{pmatrix} -1 \end{pmatrix}) = \{ z \in \setZ \setbar 1 \leq z \leq 3 \}$, corresponding to the three points on the line inside the rectangle.
}
\label{fig:deriv_lattice1}
\end{figure}
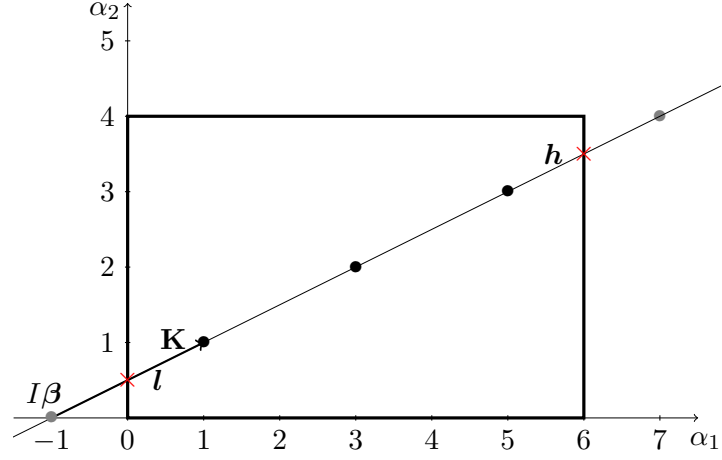

Since we have the convex set
\begin{align}
\Gamma(\vec{\beta}^p) 
&= \{I \vec{\beta}^p + K \vec{z} \setbar C \vec{\beta}^p = \vec{0} \,\land\, I \vec{\beta}^p \in \setZ^{D_f} \,\land\, \vec{z} \in \Sigma(\vec{\beta}^p) \} \nonumber \\
&= \{I \vec{\beta}^p + K \vec{z} \setbar C \vec{\beta}^p = \vec{0} \,\land\, \vec{z} \in \Sigma(\vec{\beta}^p) \} \,,
\end{align}
where we were able to drop the integer condition on $I \vec{\beta}^p$, because it is redundant to $\Sigma(\vec{\beta}^p)$ being not empty, we can now expand the sum in \cref{eq:ewd_dxp_beta_2} and thus write down an explicit expression for $\d x^p_{\vec{\beta}^p}$.
This gives
\begin{align}
\d x^p_{\vec{\beta}^p} = 
\delta_{C\vec{\beta}^p}
& \sum_{z_\kappa = \ceil{\max (L^\kappa \vec{b})}}^{\floor{\min (H^\kappa \vec{b})}} \,\,\,\,
\sum_{z_{\kappa-1} = \ceil{\max (L^{\kappa-1} \vec{b} + \widehat{L}^{\kappa-1} \vec{z}_{\kappa})}}^{\floor{\min (H^{\kappa-1} \vec{b} + \widehat{H}^{\kappa-1} \vec{z}_{\kappa})}} \cdots \nonumber \\
& \sum_{z_1 = \ceil{\max (L^{1} \vec{b} + \widehat{L}^{1} \vec{z}_{2 \,\dots\, \kappa})}}^{\floor{\min (H^{1} \vec{b} + \widehat{H}^{1} \vec{z}_{2 \,\dots\, \kappa})}}   
\d f_{I \vec{\beta}^p + K \vec{z}} \, \left. \frac{\partial \fbar}{\partial x^p_{A^p \vec{\alpha}}} \right|_{\vec{\alpha} = I \vec{\beta}^p + K \vec{z}} \,, \label{eq:ewd_dxp_sum}
\end{align}
where no Kronecker delta occurs within the sums and thus all iterations are utilized.
The evaluation of the sums can be parallelized without difficulty and no synchronization is necessary for writes to $\d x^p_{\vec{\beta}^p}$ since in this form one thread can be used per element of $\d x^p$.

\subsection{Handling expressions containing sums}\label{sec:impl_ewd_sum}
As mentioned earlier we also want to handle expressions that contain summations over one or more indices.
For this purpose consider a function containing a summation depending on arguments $y^1, \dots, y^{P'}$.
It can be written in the form
\begin{equation}
f_{\vec{\alpha}} (y^1, \dots, y^{P'}, x^1, \dots, x^P) = \fbar \!\left(\overline{s}(y^1, \dots, y^{P'}), x^1_{A^1 \vec{\alpha}}, \dots, x^P_{A^P \vec{\alpha}} \right) \,
\end{equation}
with
\begin{equation}
\overline{s}(y^1, \dots, y^{P'}) = \sum_{k \in \Psi} s(y^1_{\widehat{A}^1 \vec{\alphahat}}, \dots, y^{P'}_{\widehat{A}^{P'} \vec{\alphahat}})
\end{equation}
where $s: \setR \times \cdots \times \setR \to \setR$ and $\vec{\alphahat} \teq \begin{bmatrix} \vec{\alpha} & k \end{bmatrix}$ and $\widehat{A}^{p'}: \setZ^{D_f+1} \to \setZ^{D_{p'}}$ and $\Psi \subset \setZ$ is a convex integer set.
Using the chain rule to calculate the derivative of $l$ (defined as before) \wrt $y^{p'}_{\vec{\beta}^{p'}}$ gives
\begin{align}
\d y^{p'}_{\vec{\beta}^{p'}} &= \frac{\partial l}{\partial y^{p'}_{\vec{\beta}^{p'}}} = \sum_{\vec{1} \leq \vec{\alpha} \leq \vec{N}^f} \frac{\partial l}{\partial f_{\vec{\alpha}}} \frac{\partial f_{\vec{\alpha}}}{\partial y^{p'}_{\vec{\beta}^{p'}}} = \sum_{\vec{1} \leq \vec{\alpha} \leq \vec{N}^f} \d f_{\vec{\alpha}} \, \frac{\partial \fbar}{\partial \overline{s}} \, \sum_{\substack{k \in \Psi \\ \widehat{A}^{p'} \vec{\alphahat} = \vec{\beta}^{p'}}} \frac{\partial s}{\partial y^{p'}_{\widehat{A}^{p'} \vec{\alphahat}}} \nonumber \\
&= \sum_{\substack{\vec{1} \leq \vec{\alpha} \leq \vec{N}^f \\ k \in \Psi \\ \widehat{A}^{p'} \vec{\alphahat} = \vec{\beta}^{p'}}} \d f_{\vec{\alpha}} \, \frac{\partial \fbar}{\partial \overline{s}} \, \frac{\partial s}{\partial y^{p'}_{\widehat{A}^{p'} \vec{\alphahat}}} 
= \sum_{\vec{\alphahat} \in \widehat{\Gamma}} \d f_{\vec{\alpha}} \, \frac{\partial \fhat}{\partial y^{p'}_{\widehat{A}^{p'} \vec{\alphahat}}} \label{eq:ewd_dyp}
\end{align}
with the ``sum-liberated'' scalar function
\begin{equation}
\fhat(y^1_{\widehat{A}^1 \vec{\alphahat}}, \dots, y^{P'}_{\widehat{A}^{P'} \vec{\alphahat}}, x^1_{A^1\vec{\alpha}}, \dots, x^P_{A^P\vec{\alpha}}) \teq \fbar\!\left(s(y^1_{\widehat{A}^1 \vec{\alphahat}}, \dots, y^{P'}_{\widehat{A}^{P'} \vec{\alphahat}}), x^1_{A^1 \vec{\alpha}}, \dots, x^P_{A^P \vec{\alpha}} \right)
\end{equation}
and the ``sum-extended'' multi-index set
\begin{equation}
\widehat{\Gamma} \teq \left\{ \begin{bmatrix} \vec{\alpha} & k \end{bmatrix} \bigg|\, \vec{\alpha} \in \setZ^{D_f} \,\land\, k \in \Psi \,\land\, \vec{1} \leq \vec{\alpha} \leq \vec{N}^f \,\land\, \widehat{A}^{p'} \begin{bmatrix} \vec{\alpha} & k \end{bmatrix} = \vec{\beta}^{p'}  \right\} \,.
\end{equation}
Note that \eqref{eq:ewd_dyp} equals the original expression for the derivative \eqref{eq:ewd_dxp_beta_2} but with $\fbar$ replaced by $\fhat$, which is the same as $\fbar$ but with the sum symbol removed, and $\Gamma$ replaced by $\widehat{\Gamma}$, which additionally includes the conditions on $k$ from the sum.

Thus handling summations can be done using the previously described strategy for derivation by extending it as follows.
Each sum symbol (!) in the function $f$ to be derived is removed, its summation index is appended to the multi-index $\vec{\alpha}$ of $f$ and its summation range is included as an additional constraint in the set $\Gamma$.
This process is iterated for nested sums.
When indexing into $\d f$ the additional indices in $\vec{\alpha}$ introduced by sums are ignored.

\subsection{Element-wise Derivation Algorithm}
\Cref{alg:elemdiff} computes expressions for derivatives $\d x^p_{\vec{\beta}^p} = \partial l / \partial x^p_{\vec{\beta}^p}$ of a element-wise defined function $f$.
If an expression for the Jacobian $\partial f_{\vec{\alpha'}} / \partial x^p_{\vec{\beta}^p}$ is desired, it can be obtained from $\d x^p_{\vec{\beta}^p}$ by substituting
\[ \d f_{\vec{\alpha}} \teq \prod_d \delta_{\alpha_d,\alpha'_d}  \,. \]

Since the summation ranges \eqref{eq:ewd_sigma2} in a produced derivative are of the same form as the index ranges \eqref{eq:ewd_sigma} of the input function and we shown in \cref{sec:impl_ewd_sum} how to handle summations in the input function, we can iteratively apply the derivation algorithm on derivative functions to obtain second and higher order derivatives.
Therefore the set of element-wise defined functions using linear combination of indices for its arguments is closed under the operation of derivation.

\begin{algorithm}[tbp]\DontPrintSemicolon
\caption{Element-wise expression derivation}\label{alg:elemdiff}
\KwIn{element-wise defined tensor-valued function $f$ taking $P$ tensor arguments $x^1, \dots, x^P$; expression of derivative $\d f_{\vec{\alpha}} \teq \partial l / \partial f_{\vec{\alpha}} $}
\KwOut{expression of derivatives \wrt arguments $\d x^p_{\vec{\beta}^p} \teq \partial l / \partial x^p_{\vec{\beta}^p}$ }
\vspace{2mm}
\For(\tcp*[f]{loop over arguments $x^p$}){$p \in \{1, \dots, P\}$}{		
	$\d x^p_{\vec{\beta}^p} \longleftarrow 0$ \\
	\For(\tcp*[f]{loop over index expressions for $x^p$}){$q \in \{1, \dots, Q_p\}$}{	
		\tcp{compute derivative expression \wrt $x^p_{A^{pq}\vec{\alpha}}$ using reverse accumulation automatic differentiation (sec.~\ref{sec:impl_autodiff_graph})} 
		$\Delta \longleftarrow \d f_{\vec{\alpha}} \, \frac{\partial f_{\vec{\alpha}} }{\partial x^p_{A^{pq} \vec{\alpha}}}$ \tcp*{ignore sum symbols within $f$}
		\tcp{compute range constraints from shape of $f$ and limits of occurring sums}
		$\Omega \longleftarrow \{\text{range constraints on $\vec{\alpha}$ of the form } R \vec{\alpha} \geq r \}$ \\
		\vspace{2mm}
		\tcp{compute Smith normal form (sec.~\ref{sec:integer_eqs}) to obtain the following}
		$I \longleftarrow $integer pseudo-inverse of $A^{pq}$ \\
		$K \longleftarrow $integer kernel of $A^{pq}$ \\		
		$C \longleftarrow $integer cokernel of $A^{pq}$ \\		
		\vspace{2mm}
		\tcp{rewrite constraints using $\vec{\beta}^p$ and kernel factors $\vec{z}$}
		$\Omega' \longleftarrow \{R \, K \vec{z} \geq r - R \, I \vec{\beta}^p \setbar (R \vec{\alpha} \geq r) \in \Omega \}$ \\
		\tcp{solve $\Omega'$ for $\vec{z}$ using Fourier-Motzkin elimination (sec.~\ref{sec:lin_ieq})}
		$\Sigma \longleftarrow \{\text{range constraints }\Omega' \text{ on }\vec{z}\text{ transformed into form \eqref{eq:ewd_sigma2}}\}$ \\
		\vspace{2mm}
		\tcp{generate derivative expressions}
		$\displaystyle \d x^p_{\vec{\beta}^p} \longleftarrow \d x^p_{\vec{\beta}^p} + \delta_{C\vec{\beta}^p} \sum_{\vec{z} \in \Sigma} \, \left. \Delta \right|_{\vec{\alpha} = I \vec{\beta}^p + K \vec{z}}$ \tcp*[f]{use form \eqref{eq:ewd_dxp_sum} for sum}
	}
}
\end{algorithm}

\clearpage
\section{Example and Numeric Verification}\label{sec:example}
The implementation code is provided at \url{https://github.com/surban/TensorAlgDiff}.
In our implementation and thus in this example we use zero-based indexing, \ie a vector $x \in \setR^{N}$ has indices $\{0, \dots, N-1\}$, as it is usual in modern programming languages.
Given the function
\[ f_{ij}(a,b,c,\vec{d}) = \exp\left[- \sum_{k=0}^4 \left( (a_{ik} + b_{jk})^2 \, c_{ii} + d_{i+k}^3 \right)  \right] \]
where the shapes of the arguments are $a \in \setR^{3 \times 5}$, $b \in \setR^{4 \times 5}$, $c \in \setR^{3 \times 3}$ and $d \in \setR^{8}$ and the shape of $f$ is $f \in \setR^{3 \times 4}$ the derivation algorithm produces the following output:
\begin{lstlisting}[breaklines]
Input: f[i; j] = exp (-sum{k}_0^4 (((a[i; k] + b[j; k]) ** 2 * c[i; i] + d[i + k] ** 3)))
Derivative of f wrt. a: da[da_0; da_1] = sum{da_z0}_0^3 (((-(df[da_0; da_z0] * exp (-sum{k}_0^4 (((a[da_0; k] + b[da_z0; k]) ** 2 * c[da_0; da_0] + d[da_0 + k] ** 3))))) * c[da_0; da_0] * 2 * (a[da_0; da_1] + b[da_z0; da_1]) ** (2 - 1)))
Derivative of f wrt. b: db[db_0; db_1] = sum{db_z0}_0^2 (((-(df[db_z0; db_0] * exp (-sum{k}_0^4 (((a[db_z0; k] + b[db_0; k]) ** 2 * c[db_z0; db_z0] + d[db_z0 + k] ** 3))))) * c[db_z0; db_z0] * 2 * (a[db_z0; db_1] + b[db_0; db_1]) ** (2 - 1)))
Derivative of f wrt. c: dc[dc_0; dc_1] = if {dc_0 + -dc_1 = 0} then (sum{dc_z1}_0^4 (sum{dc_z0}_0^3 (((a[dc_1; dc_z1] + b[dc_z0; dc_z1]) ** 2 * (-(df[dc_1; dc_z0] * exp (-sum{k}_0^4 (((a[dc_1; k] + b[dc_z0; k]) ** 2 * c[dc_1; dc_1] + d[dc_1 + k] ** 3))))))))) else (0)
Derivative of f wrt. d: dd[dd_0] = sum{dd_z1}_(max [0; -2 + dd_0])^(min [4; dd_0]) (sum{dd_z0}_0^3 (((-(df[dd_0 + -dd_z1; dd_z0] * exp (-sum{k}_0^4 (((a[dd_0 + -dd_z1; k] + b[dd_z0; k]) ** 2 * c[dd_0 + -dd_z1; dd_0 + -dd_z1] + d[dd_0 + -dd_z1 + k] ** 3))))) * 3 * d[dd_0] ** (3 - 1))))
\end{lstlisting}
The operator \verb|**| denotes exponentiation in this output.
The Kronecker delta has been encoded as a ``\emph{if} $x$ \emph{then} $y$ \emph{else} $z$'' expression for more efficiency.
Internally these expressions are represented as graphs, thus subexpressions occurring multiple times are only stored and evaluated once and no expression blowup as with symbolic differentiation occurs.
To cleanup the generated expressions from the automatic differentiation algorithm an expression optimization step, which pre-evaluates constant parts of the expressions, should be incorporated.
However, since this is not part of the core derivation problem, it has not been performed for this demonstration.
These derivative expressions have been verified by using random numeric values for the arguments and comparing the resulting values for the Jacobians with results from numeric differentiation.

\newpage
\section{Conclusion}
We have presented a method to compute symbolic expressions for derivatives of element-wise defined tensor-valued functions.
These functions may contain summations and the indices of its arguments can be an arbitrary linear combination of the function indices.
The output of our algorithm is an explicit symbolic expression for each element of the derivative.
Thus the resulting expressions are very well suited for massively parallel evaluation in a lock- and synchronization-free CUDA kernel, which computes one element of the derivative per thread.
No temporary memory is necessary for the evaluation of the derivatives.
The derivatives themselves may contain additional summations over indices which have become free in the derivative.
The output of the algorithm specifies the ranges of these sums as a maximum or minimum over a set of linear combinations of the derivative indices; therefore computing the numerical range at evaluation time costs only two matrix multiplications per loop run (not iteration).

\newpage
\bibliographystyle{apalike}
\bibliography{elemdiff}

\end{document}